\documentclass[prx,epsf,showpacs,twocolumn,superscriptaddress,footinbib]{revtex4-2}

\usepackage[pdftex]{graphicx}
\usepackage{dcolumn}
\usepackage{bm}
\usepackage{epsfig}
\usepackage{latexsym}
\usepackage{amsmath}
\usepackage{amssymb}
\usepackage{amsfonts}
\usepackage{physics}
\usepackage[position=top,caption=false,singlelinecheck=off, justification=raggedright]{subfig}
\usepackage{color}
\usepackage{array}
\usepackage{bbold}
\usepackage{dsfont}
\usepackage{mathrsfs}
\usepackage{stmaryrd}
\usepackage{listings}
\usepackage{tikz}
\usetikzlibrary{quantikz2}
\usepackage{hyperref}
\usepackage{cleveref}
\hypersetup{
    colorlinks,
    citecolor=blue,
    filecolor=red,
    linkcolor=magenta,
    urlcolor=blue
}

\usepackage{xr}
\makeatletter
\newcommand*{\addFileDependency}[1]{
  \typeout{(#1)}
  \@addtofilelist{#1}
  \IfFileExists{#1}{}{\typeout{No file #1.}}
}
\makeatother

\newcommand*{\myexternaldocument}[1]{%
    \externaldocument{#1}%
    \addFileDependency{#1.tex}%
    \addFileDependency{#1.aux}%
}
\myexternaldocument{supplement}

\usepackage{algpseudocode}

\newcommand{\cnot}{CNOT}

\newcommand{\tgate}{T}
\newcommand{\sgate}{S}

\newcommand{\nkd}[3]{\left \llbracket #1,\, #2,\, #3 \right \rrbracket}

\DeclarePairedDelimiter\floor{\lfloor}{\rfloor}

\usepackage[normalem]{ulem}

\usepackage[dvipsnames]{xcolor}
\usepackage{hyperref}
\hypersetup{
  colorlinks   = true,
  urlcolor     = Blue,
  linkcolor    = BrickRed,
  citecolor   = PineGreen
}

\newcommand{\chris}[1]{\textcolor{black}{#1}}

\def\cambridge{Quantinuum, Terrington House, Cambridge, CB2 1NL, UK}
\def\london{Quantinuum, Partnership House, Carlisle Place, London SW1P 1BX, UK}

\usepackage{graphicx}
\usepackage{dcolumn}
\usepackage{tabularx}
\usepackage{xcolor}
\usepackage{bm}

\usepackage{centernot}
\begin{document}

\title{Fast verification of fault tolerant magic state preparation circuits \\ with the CNOT + T gate set}

\author{Selwyn Simsek}
\affiliation{\london}
\author{Callum Macpherson}
\affiliation{\cambridge}
\author{Chris N. Self}
\affiliation{\cambridge}
\author{David Amaro}
\affiliation{\london}

\date{\today}

\begin{abstract}
Magic state preparation is often among the most expensive operations in a fault tolerant quantum computation. 
One obstacle to coming up with better magic state preparation circuits lies in verifying their fault tolerance properties, as it is difficult to verify either by hand or in software that a very large number of combinations of faults do not give rise to high-weight or logical errors, especially in a non-Clifford circuit.
In this work, we propose the use of the CNOT + T gate set to fault-tolerantly prepare a large class of magic states.
We consider the propagation of Pauli faults within CNOT + T circuits and derive properties similar to those of Pauli faults within Clifford circuits, with a view to verifying fault tolerance as efficiently as possible.
We present Comb, a library that can verify the fault tolerance of CNOT + T magic state preparation circuits, and demonstrate its use in verifying the fault tolerance of distance three and distance five magic state preparation circuits in minutes on a laptop.
We present a distance five magic state preparation circuit that was discovered using Comb.
As an independent confirmation of Comb's correctness we compare its findings to circuit-level noisy simulations for several magic state preparation circuits.
\end{abstract}
\maketitle
\section{Introduction} \label{sec:introduction}

\begin{figure*}[t]
\centering
\includegraphics[width=\textwidth]{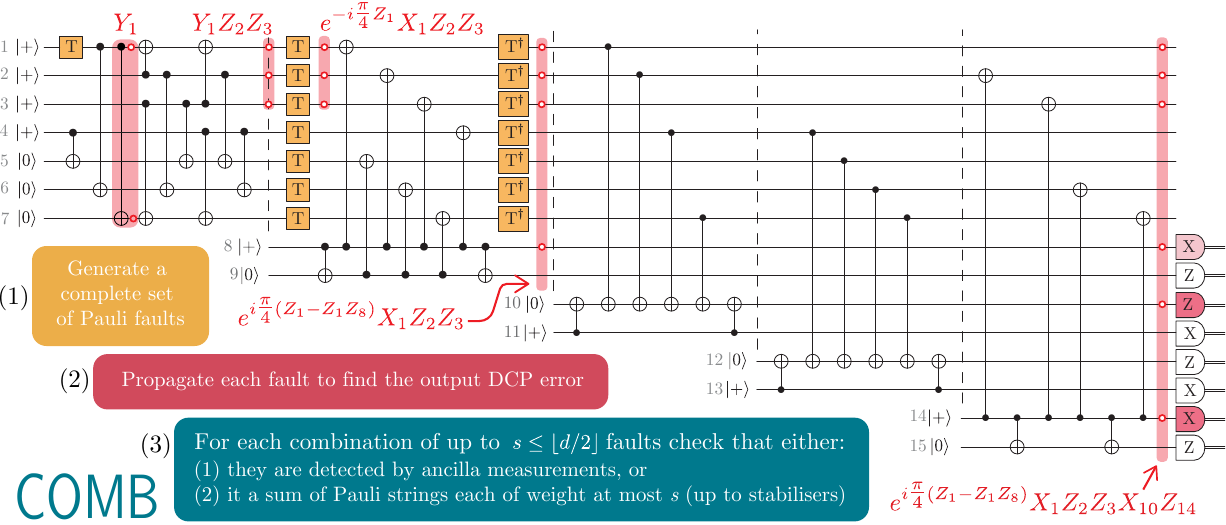}
\caption{
Workflow for checking the fault-tolerance of a CNOT + T circuit. (1) We begin by generating a complete set of Pauli circuit faults. (2) Each Pauli fault is propagated through the circuit. In general this will produce errors that are linear combinations of Paulis, however because of the structure of the CNOT + T gate set these errors remain efficiently describable as \emph{Diagonal Clifford and Pauli} (DCP) errors. We show an example of propagating a Pauli fault into a DCP error in a CNOT + T circuit for fault-tolerantly preparing $T|+\rangle$ in the $\nkd{7}{1}{3}$ code. 
Post-selecting on flag measurements detects any Pauli errors that anti-commute with the measurement basis. However, for DCP errors some of their Pauli components can anti-commute with the measurement basis while others commute with it and so can persist after post-selection. 
This situation is illustrated here: post-selecting on the measurements of either qubits 10 and 14 would detect this error, however if we only post-selected on qubit 8 part of the error would not be caught.
(3) To check the circuit is fault-tolerant Comb considers all combinations of $s \leq \lfloor d/2 \rfloor$ faults, where $d$ is the distance of the code, and checks that none of the propagated errors contain an undetected Pauli string of weight greater than $s$. By carefully exploiting the structure of DCP errors we avoid the exponential memory cost of naively expanding them into Pauli strings.
\label{fig:main-fig}}
\end{figure*}

Fault-tolerant quantum computation is now an experimental reality with much recent work demonstrating some or all of the required ingredients to perform a universal computation, such as state preparation \cite{ryan-andersonRealizationRealTimeFaultTolerant2021b}, single qubit gates \cite{ryan-andersonRealizationRealTimeFaultTolerant2021b,dasuBreakingEvenMagic2025} and entangling gates \cite{ryan-andersonImplementingFaulttolerantEntangling2022}. However, since the Eastin-Knill theorem \cite{eastinRestrictionsTransversalEncoded2009} states that it is not possible to carry out a universal gate set transversally, some of these operations must consume considerable quantum resources, which leads to a performance bottleneck in fault tolerant quantum computing.

One popular approach is to choose a code in which the Clifford group can be implemented transversally, and carry out a single non-Clifford gate such as T or CCZ by injecting a magic state into the code, which is sufficient for universality \cite{bravyiUniversalQuantumComputation2005}. When magic states \cite{bravyiUniversalQuantumComputation2005} were introduced, it was proposed to prepare magic states fault tolerantly by means of magic state distillation, in which multiple noisy copies of magic states are prepared and then `distilled' to produce a single magic state with a lower error rate.
Since the introduction of magic state distillation~\cite{bravyiUniversalQuantumComputation2005}, many improvements and alternative methods have been proposed \cite{campbellUnifiedFrameworkMagic2017,  litinskiGameSurfaceCodes2019, gidneyEfficientMagicState2019}, leading to lower estimates of the resources that these methods consume, in some cases giving improvements in the resources consumed of orders of magnitude.
These results have greatly improved the outlook of fault tolerant quantum computing.

However, these methods mostly concern magic state preparation in the asymptotic regime, often limited to the surface code.
In the near term, one is often interested in fixed codes of small size, such as small instances of the colour code \cite{bombinIntroductionTopologicalQuantum2013}, as these have high rates compared to surface codes of similar size, allowing for more interesting demonstrations to be performed with them on contemporary quantum computers \cite{ryan-andersonRealizationRealTimeFaultTolerant2021b, ryan-andersonImplementingFaulttolerantEntangling2022}.
With such codes, the prospect exists of preparing the magic states directly without distillation, which is to say without the requirement of preparing several magic states at once.
Preparing magic states this way naturally reduces the spatial overhead associated with magic state preparation \cite{chamberlandFaulttolerantMagicState2019, chamberlandVeryLowOverhead2020a,gidneyMagicStateCultivation2024} and, in some cases, makes the number of gates required comparable to that of syndrome extraction \cite{gotoMinimizingResourceOverheads2016}.

It is challenging to construct circuits that prepare magic states fault tolerantly, owing to the fact that within non-Clifford circuits, Pauli faults do not necessarily result in output errors that are also Pauli, which leads to a more complicated analysis than is the case for Clifford circuits \cite{surtiEfficientSimulationLogical2025}.

One popular approach is to `Cliffordise' the circuit, which is to say replacing T gates with S gates \cite{gidneyCleanerMagicStates2023,gidneyMagicStateCultivation2024,surtiEfficientSimulationLogical2025}. While this can lead to a great reduction in computational effort when simulating, it can potentially lead us to miss problematic faults and come to the wrong conclusion about a circuit's fault tolerance, as we discuss in \Cref{sec:appendix-counterexample}.

Additionally, it is hard to reason analytically and computationally about the fault tolerance of circuits beyond distance three owing to the potentially very large number of combinations of faults in a circuit, each of which must be checked so that it does not cause an error that violates the requirements of fault tolerance.

In this work, we aim to make this task easier by noting that the CNOT + T gate set is capable of preparing a wide range of magic states fault-tolerantly, and showing that it is easier to verify the fault tolerance of circuits that are restricted to the CNOT + T gate set than those that are Clifford + T. We develop a library called Comb that verifies the fault tolerance of a magic state preparation circuit by iterating over the set of combinations of faults that appear in such circuits as efficiently as possible. 
We illustrate the workflow of checking the fault-tolerance of a CNOT + T circuit in \Cref{fig:main-fig}.
By verifying the fault tolerance of a magic state preparation circuit quickly in software, we aim to facilitate the iterative design of such circuits, which is important as few (if any) analytic or software tools satisfying a similar function exist for any gate set that is not Clifford.

In \Cref{sec:preliminaries}, we will define some mathematical preliminaries.
In \Cref{sec:cnot-plus-t}, we review the properties of the CNOT + T gate set with a view to preparing magic states fault-tolerantly, and discuss how faults propagate through circuits composed of this gate set.
In \Cref{sec:verifying-fault-tolerance}, we discuss in detail how to computationally verify the fault tolerance of CNOT + T circuits.
In \Cref{sec:demonstration}, we demonstrate the performance of Comb, a library that implements the results of \Cref{sec:verifying-fault-tolerance}, on a test suite of circuits, and show that it is feasible to verify the fault tolerance of magic state preparation circuits of reasonable size, including a novel circuit that was discovered using Comb which prepares a magic state fault-tolerantly to distance five. Additionally, we simulate some of the test suite circuits with circuit-level noise and check that these simulations are in agreement with Comb.

\begin{figure*} 
\centering
\includegraphics[width=0.95\textwidth]{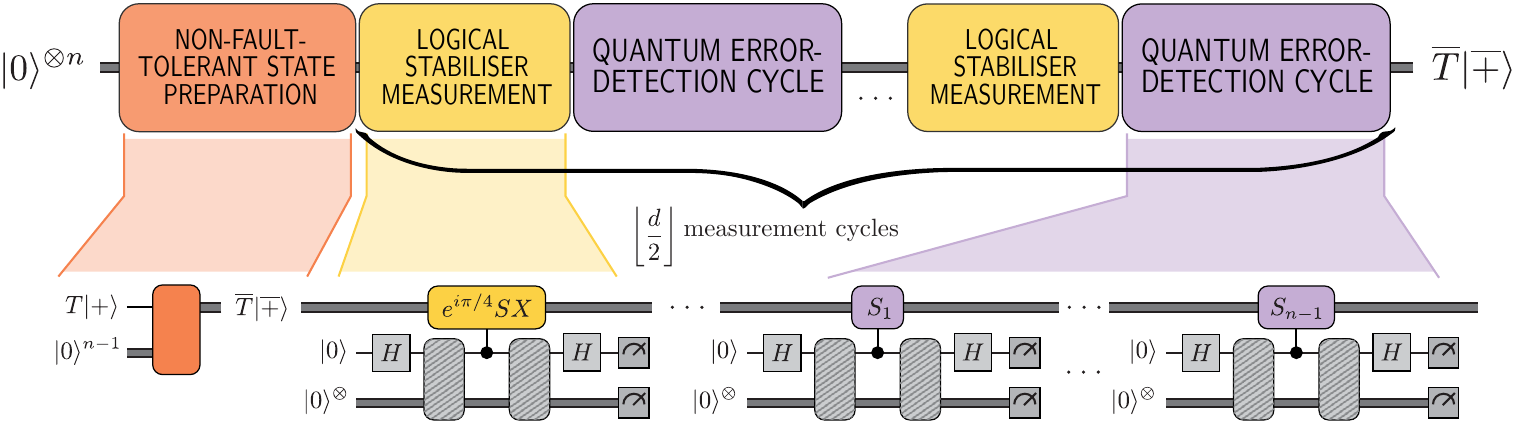}
\caption{Schematic illustration of a fault-tolerant $\overline{T}|\overline{+}\rangle$ magic state preparation circuit in the CNOT + T gate set for an $\nkd{n}{1}{d}$ code, based on the construction of~\cite{chamberlandFaulttolerantMagicState2019,chamberlandVeryLowOverhead2020a}. An initial encoder circuit non-fault-tolerantly encodes the physical $T|+\rangle$ state into the code space. This is followed by $\lfloor d/2 \rfloor$ rounds of measurements of the logical stabiliser $e^{-i\pi/4} SX$ and the stabiliser generators of the code $S_i$. We call this operator the logical stabiliser in the sense that $(e^{-i\pi/4} SX) \, T |+\rangle = T |+\rangle$. The outcomes of these measurements will be post-selected on, meaning if any of them return a $(-1)$ the state preparation is rejected. To make these measurements fault-tolerant they will typically be flagged, indicated by the hatched boxes involving additional flag ancillas. Qubits are initialised and measured in either the $X$ or $Z$ basis, all other gates are CNOT or T.} \label{fig:chamberland-schematic}
\end{figure*}

\section{Preliminaries} \label{sec:preliminaries}

We will consider the Pauli group that contains the Pauli matrices $X,Y,Z$ together with phase,
\begin{equation}
\mathcal{P}=\langle i,X,Z \rangle,
\end{equation}
and the Pauli group on $n$ qubits $\mathcal{P}_n$ to be the $n$-fold product
\begin{equation}
\mathcal{P}_n=\langle i,X,Z \rangle^{\otimes n}
\end{equation}
of the Pauli group. Subsequently, we will often use the term `Pauli' to refer to an element of this Pauli group $\mathcal{P}_n$ on $n$ qubits, and not just elements of the single qubit group $\mathcal{P}$.
We will refer to Paulis as phaseless or lacking phase if they do not possess a leading non-trivial phase when written out as products of $X$, $Y$, $Z$ and phase. For example, the Pauli $XZ$ is phaseless while the Pauli $iXY$ is not.
We also define Paulis of $X$ type, or $X$ Paulis for short, to be those that are products of the identity or the Pauli $X$, together with phase,
\begin{equation}
\mathcal{P}^X_n=\langle i,X \rangle^{\otimes n}.
\end{equation}
We define Paulis of $Z$ type similarly.

Since $Y = i XZ$, we note that it is always possible to write a Pauli $P$ as a product of a Pauli of $X$ type $P^X$ and a Pauli of $Z$ type $P^Z$,
\begin{equation}
P=P^XP^Z, \label{preliminaries:pauli-decomposition}
\end{equation}
a decomposition that we will sometimes make use of. It is possible to assume that either $P^X$ or $P^Z$ is phaseless, or reverse their order in the product, without loss of generality.
We will quotient out phase in defining the Clifford group $\mathcal{C}^n$,
\begin{equation}
\mathcal{C}_n = \{U: U^\dagger \mathcal{P}_n U = \mathcal P_n \} / \left \{ e^{i \theta} \mathbb{1}_n : \theta \in \mathbb{R} \right \},
\end{equation}
so that in particular it is finite, and elements of it can be stored in memory on a computer.
We refer the interested reader to \cite{dehaeneCliffordGroupStabilizer2003} for a treatment of how to store and manipulate Paulis and Cliffords computationally.

We will also define the $S=\text{diag}(1,i)$ and $T=\text{diag}(1,e^{\frac{i \pi}{4}})$ gates to avoid confusion with definitions in other works. We note that $S=T^2$ and $T^\dagger=T^7$, meaning that $S$, $T$ and their inverses are all generated by $T$.

We will also make use of the CNOT gate, which is
\begin{equation}
    \cnot= \ket{0}\bra{0} \otimes \mathbb{1} + \ket{1}\bra{1}\otimes X,
\end{equation}
adding subscripts when appropriate to make clear what the indices of the control and target qubits are.
When we refer to measurements, they will be assumed to be Pauli measurements in either the $X$ or the $Z$ basis, making clear which one is intended when necessary.

\section{The CNOT + T gate set}  \label{sec:cnot-plus-t}

\subsection{Applicability}

The CNOT + T gate set is already widely used to prepare magic states fault-tolerantly though this fact is rarely noted explicitly. Example are the flagged fault-tolerant construction of~\cite{chamberlandFaulttolerantMagicState2019}, magic state cultivation \cite{gidneyMagicStateCultivation2024} and distillation \cite{fowlerLowOverheadQuantum2019}. It also has been studied in the context of reducing the T count of a logical circuit \cite{heyfronEfficientQuantumCompiler2018}, and polynomial-time simulation \cite{cercelescuPolynomialtimeSimulationNonClifford2026}.

An illustration of fault tolerantly preparing a magic states in the CNOT + T gate set is shown schematically in \Cref{fig:chamberland-schematic}, based on the flagged fault-tolerant construction of~\cite{chamberlandFaulttolerantMagicState2019}.
The sketched circuit prepares a magic state $T\ket+$ in an $\nkd{n}{1}{d}$ code. In this circuit, as in all CNOT + T circuits that we consider in this paper, qubits are initialised in either the $X$ or $Z$ eigenstates $\ket 0$ and $\ket +$ and any measurements that are postselected on are made in the $X$ or $Z$ bases.

\subsection{DCP operators} \label{sec:dcp-operators}
We now state and prove some results concerning the propagation of faults through magic state preparation circuits, which will be useful in verifying the fault tolerance of such circuits.

The CNOT + T gate set preserves a set of operators under conjugation. We define a {\it Diagonal Clifford and Pauli} (DCP) operator $C_{\text{DCP}}$ to be the product of a diagonal Clifford $C_\text{D}$ and a Pauli $P$,
\begin{equation}
    C_{\text{DCP}} = C_\text{D} P.
\end{equation}
We will assume, without loss of generality, that the Pauli $P$ is of $X$ type, as any $Z$ Pauli can be absorbed into the definition of the diagonal Clifford $C_\text{D}$.
\chris{
We can write the diagonal Clifford $C_\text{D}$ as the exponential of a linear combination of Paulis of Z type $D_i$, which are diagonal, so that
\begin{equation}
C_{\text{DCP}} = \exp\left(\frac{i \pi}{4}\sum_{l=1}^r D_l\right) P.
\label{eqn:DCP-with-exp}
\end{equation}
Expanding the exponential would yield an expression for the DCP error as a sum of Pauli strings.
}

Under conjugation by CNOT or T, any DCP operator is mapped to another DCP operator. 
\chris{Using Equation~\eqref{eqn:DCP-with-exp} we see that conjugating $C_{\text{DCP}}$ by $\cnot_{i,j}$ to obtain $C'_{\text{DCP}}$ gives}
\begin{align}
C_{\text{DCP}}' =&\ \exp\left(\frac{i \pi}{4}\sum_{l=1}^r \cnot_{i,j} \,D_l \,\cnot_{i,j}\right) \nonumber \\ &\ \cdot\cnot_{i,j}\, P \,\cnot_{i,j}.
\end{align}
As the CNOT gate preserves the $X$ or $Z$ type of a Pauli under conjugation, the terms $\cnot_{i,j} \,D_l\, \cnot_{i,j}$ that appear in the sum in the exponential are also Paulis of $Z$ type. Therefore, the resulting operator $C'_{\text{DCP}}$ is also DCP.

Now, conjugating $C_{\text{DCP}}$ by a T gate $\tgate_i$ to obtain $C'_{\text{DCP}}$ gives that
\begin{equation}
C'_{\text{DCP}} = \exp\left(\frac{i \pi}{4}\sum_{l=1}^r D_l\right) \tgate_i\, P \,\tgate_i^\dagger,
\end{equation}
where we have used the fact that diagonal operators commute in order to push the $\tgate_i$ gate through the exponential.
If the Pauli $P$ does not have support on the $i$-th qubit, then it will commute with $\tgate_i$ and so $C_{\text{DCP}}=C'_{\text{DCP}}$, meaning that $C'_{\text{DCP}}$ is also DCP. If the Pauli $P$ does have support on the $i$-th qubit, then
\begin{equation}
\tgate_i \,P \,\tgate_i^\dagger = e^{-\frac{i \pi}{4}} \text{S}_i P,
\end{equation}
which again means that $C'_{\text{DCP}}$ is DCP, as $e^{-\frac{i \pi}{4}} \sgate_i$ is a diagonal Clifford. Since CNOT and T gates take DCP operators to DCP operators under conjugation, so will any product of them.

The DCP operators of fixed size $n$ form a group, and they contain the $n$-qubit Paulis $\mathcal{P}_n$ as a subgroup. They play the same role with respect to the CNOT + T gate set that the Pauli group does for the Clifford gate set, as both are preserved under conjugation by the respective gate set. Since Paulis are DCP, and CNOT + T preserves DCP operators, it follows that Pauli faults in a CNOT + T circuit will propagate to output errors which are DCP, which hints at the practical importance of the DCP group for fault tolerant analysis.

However, for the purposes of verifying fault tolerance, it is not the group-theoretic properties of the DCP group that are most relevant, but rather the representational properties of the group. As the DCP group lies inside the Clifford group, it can be represented up to phase in matrix form using $\mathcal{O}(n^2)$ bits \cite{dehaeneCliffordGroupStabilizer2003}, and multiplication of DCP operators is roughly equivalent to multiplication of these matrices.

This is not much less efficient than the Pauli group, for which Paulis are often represented up to phase with $2n$ bits, and multiplication of Paulis is equivalent to addition mod 2, which is of complexity $\mathcal{O}(n)$. The efficient storage and manipulation of DCP operators will also prove useful for fault tolerant analysis.

The discussion so far neglects the role of phase. If one wanted to take into account the phases that result from multiplying Cliffords, this representation would not suffice, and the analysis would be more complicated \cite{heinrichStabiliserTechniquesTheir2021}. However, when analysing the fault tolerance of a circuit, the phase of an output error can be neglected altogether as it cannot change the weight of the error or whether it is detected, so we will settle for representing DCP operators in the manner discussed.

Finally, we note that DCP operators are an example of Pauli-Square-Root Clifford (PSC) operators, also introduced in the context of magic state preparation \cite{surtiEfficientSimulationLogical2025}. We note that DCP operators naturally arise by considering Paulis propagating within CNOT + T circuits, whereas the more general class of PSC operators results from propagating Paulis through Clifford + T circuits that satisfy some technical conditions; for example, the only controlled-$U$ gates that can be performed must have that $U^2=\mathbb{1}$ \cite{surtiEfficientSimulationLogical2025}. We posit that our working within the CNOT + T gate set precludes the need to explicitly derive and work around constraints on the circuits we must consider, while retaining sufficient expressive power to prepare magic states for many codes.

\section{Verifying fault tolerance} \label{sec:verifying-fault-tolerance}

\subsection{Definition of fault tolerance}
\label{sec:verifying-fault-tolerance:defn}

We refer to fault-tolerance in the sense given by Gottesman \cite{gottesmanSurvivingQuantumComputer2024}. We decompose a circuit that prepares some state in terms of its {\it circuit components}. Any of these circuit components, which include bare wires, measurements, state preparation, and unitaries may fail, in which case a {\it fault} may occur immediately after the circuit component on the qubits where it acts. A combination of faults will cause some corresponding {\it error} at the output.

A CNOT + T state preparation circuit is fault-tolerant to distance $d$ if it obeys the condition:
\begin{enumerate}
    \item[($\star$)] any combination of $s  \leq t=\floor{\frac{d}{2}}$ faults produces an output error that is either detected by a pattern of unexpected ancilla measurements or is a sum of Pauli strings each of weight at most $s$.
\end{enumerate}
The weight of the Pauli strings is taken modulo multiplication by the stabiliser group of the code that the state belongs to. This is because multiplying an error by a stabiliser gives an equivalent error which has precisely the same effect on the state. Consequently, an error that appears to be of high weight may be equivalent to one of weight less than or equal to $s$ upon multiplication by a stabiliser, and such errors are not problematic from the perspective of fault tolerance.

It is common to refer to combinations of faults, or the corresponding output errors that they give rise to, as `good' if they satisfy this condition and `bad' otherwise. Only if all fault combinations are good can a circuit be fault-tolerant, and the existence of a single bad fault combination is sufficient to break fault tolerance.

The condition ($\star$) extends the typical definition of fault-tolerance for Pauli errors to DCP errors, whose weight is less straightforward to define.
It ensures that good output errors are still correctable, since the individual components of the error are correctable and hence by linearity the total output error is correctable. 

In Appendix~\ref{appendix:ft-condition} we show that this definition is equivalent to the more traditional criteria that appear in e.g. \cite{gottesmanSurvivingQuantumComputer2024}.

Without loss of generality, we will henceforth take all faults to be Pauli, a common assumption which has the advantage that the set of possible faults associated to each circuit component becomes finite. These Pauli faults will propagate to DCP output errors.

It is apparent from the definition of fault tolerance that the number of combinations of faults that must be considered may be very large. Assuming that there are $M$ circuit components, and that each one has at least $f$ faults associated with it, then for each $s$ one must check that at least $\binom{M}{s}f^s = \Theta(M^sf^s)$ combinations of $s$ faults from distinct locations are good. This quantity grows very quickly with $M$ even if $f$ and $s$ are not too large.

\subsection{Algorithm}
We propose to verify fault tolerance simply by checking the definition, which is to say, given a circuit, we loop over all possible combinations of $s \leq t$ faults, and verify that each error is either detected by a measurement or its weight is lower than or equal to $s$.

Although the number of fault combinations that must be considered is vast, there are many opportunities to save on computational time and space that we will take advantage of.

\subsubsection{Storing output errors}
First of all, we will address the issue of computing the output error that results from each combination of faults. Given $s$ faults $P_1,\ldots,P_s$, we write $U_1,\ldots,U_s$ for the unitaries that result from the part of the circuit subsequent to each of the $s$ faults until any postselections on ancilla qubits, which are all postponed to the end of the circuit.
The error $E_i$ at the output immediately prior to any postselected measurements that results from considering the fault $P_i$ individually is then
\begin{equation}
E_i = U_i P_i U_i^\dagger.
\end{equation}
The output error $E$ that results from the combination of the faults is equal to the product of the output errors $E_i$ that result from each fault. There is a subtlety that does not arise with Clifford circuits in that the time ordering of the faults becomes relevant. This is due to the fact that output errors are not necessarily Pauli and hence do not necessarily commute or anticommute with each other. Labelling the faults $P_1,\ldots,P_s$ so that when $i<j$, the circuit component corresponding to $P_i$ does not occur later than that of $P_j$, the combined output error $E$ is
\begin{align}
E&=U_s P_s U_s^\dagger \ldots U_1 P_1 U_1^\dagger \nonumber \\
&= E_s \ldots E_1,
\end{align}

Since $P_i$ is Pauli, from the results of \Cref{sec:dcp-operators}, the output error $E_i$ is DCP. This means that it can be evaluated efficiently. Furthermore, each output error $E_i$ that results from each fault need only be computed exactly once, and then subsequently saved for future reference. This means that computing each output error combination $E$ only involves multiplying together DCP operators that have already been computed, which, as discussed in \Cref{sec:dcp-operators}, is efficient in time and space.

We note that if we were to work in a more general gate set such as Clifford + T, then each Pauli fault $P_i$ would propagate to an output error $E_i$ that would not necessarily be Pauli or DCP, in which case storing even one output error would take space exponential in the number of qubits, let alone the cost of multiplying such operators, which illustrates one computational advantage of limiting oneself to the CNOT + T gate set.

\subsubsection{Postselecting an output error combination}
In order to check fault tolerance, we must verify whether an output error combination $E$ is good or bad. There are two parts to checking the goodness of a fault combination. We must determine whether the output error is detected, and if is not, then we must check whether the weight of the output error is below $s$ or not.

The first of these tasks, determining whether an output error is detected, is considerably more complicated than for Clifford circuits. Given a Clifford circuit, a fault $P$ results in an output error $E$ which is also Pauli. Such errors will either flip or leave invariant any ancilla measurements, depending on whether they commute or anticommute with the measurement operator. Consequently, when one postselects on the results of some ancilla measurements, a Pauli error is either always detected or always undetected, and its effect on the remaining qubits is known.

In a CNOT + T circuit, however, the errors that result from faults are DCPs, not Paulis. As a result, they do not have a deterministic effect on measurement outcomes, as DCPs do not necessarily either commute or anticommute with the measurement operators. Postselecting on ancilla measurements in the presence of a DCP error leads to a conditional error on the data qubits that may not be the same as the error prior to postselection.

We take advantage of the structure of a DCP operator to break the problem into two cases. A DCP operator $C_\text{DCP}=C_\text{D}P$ can be written as the product of a diagonal Clifford $C_\text{D}$ and a Pauli $P$ of $X$ type. Consider first any ancilla measurements that are made in the $Z$ basis. Since these measurement operators will commute with any diagonal operator, we have that a DCP error will flip the measurement deterministically depending on whether $P$ commutes or anticommutes with the measurement operator, just as is the case for Pauli errors, meaning that they can be dealt with the same way. Provided that $P$ is not detected, the conditional $X$ error $E'_X$ on the data qubits that results from postselecting on the $Z$ basis measurements will be $P$ restricted to those qubits.

We then consider postselecting on ancilla measurements that are made in the $X$ basis. The Pauli $P$ will always commute with the measurement operators since it is of $X$ type, which means that it will have no effect on the measurement outcome and can be ignored when considering postselection in the $X$ basis. As a result, we can assume for the rest of this analysis of the effect of postselection that $C_\text{DCP}=C_\text{D}$ is strictly diagonal Clifford, with no Pauli $X$ component.

In \Cref{sec:diagonal-coherent-error-analysis}, we show that, given an error $E=C_\text{D}$ which is diagonal Clifford, the conditional error $E'$ that results from postselecting on a Pauli measurement made in the $X$ basis is either another diagonal operator or vanishes (in which case the error is always detected), and can be computed efficiently. We can apply this result repeatedly to then postselect on the remaining $X$ basis ancilla measurements, and obtain the conditional diagonal error $E'_Z$ that results on the data qubits due to the $X$ basis measurements. The resulting full conditional error $E'=E'_XE'_Z$ is the product of the two conditional errors that result from considering the $X$ and $Z$ basis measurements separately, and is also DCP.

\subsubsection{Checking the weight of a conditional output error}
Now that we have the conditional output error $E'$ on the data qubits, provided that it does not vanish we must verify whether the weight of the error $E'$ is below the number of faults $s$ or not in order to decide whether the fault combination is good or not. This is made complicated by the fact that $E'$ is not necessarily Pauli. We make use of the fact that, if two errors $A$ and $B$ are correctable, then their coherent sum $A+B$ must be correctable also \cite{gottesmanSurvivingQuantumComputer2024}. We show also in \Cref{sec:diagonal-coherent-error-analysis} that a diagonal Clifford $C$ may be expressed as a linear combination of $2^r$ Paulis of $Z$ type that belong to the coset of some $r$-dimensional vector space $V=\langle P^Z_1,\ldots,P^Z_r\rangle$ whose generators $P^Z_1,\ldots,P^Z_r$ can be computed efficiently. This allows one to express the conditional output error $E'=C_\text{D}P$ as a linear combination of Paulis simply by multiplying that of the diagonal Clifford $C_\text{D}$ on the right by $P$.

We then verify that the conditional output error $E'$ is good by iterating over this sum and verifying that each element is low weight. We make use of lookup tables to compute the weight of each element in constant time, which we discuss in detail in \Cref{appendix:weight-lookup-tables}. Although each element can be checked in constant time, there are $2^r$ elements of the vector space, which is to say this operation takes time exponential in $r$.

Although it is unfortunate to finally encounter an overhead that is not polynomial in some parameter of the problem, the situation is not as bad as it appears.

First of all, there exists the opportunity to iterate over the elements of the vector space in parallel, which reduces the overhead in time. Secondly, the parameter $r$ is not necessarily that large. Without loss of generality, we are able to place the generators of the vector space in canonical form, as detailed in Appendix~\ref{sec:compression-trick}, which limits the size of $r$. For the Golay code, a code which has a higher number of qubits and distance than the other codes considered in this paper, $r$ cannot exceed 12, as also shown in Appendix~\ref{sec:compression-trick}, which is to say there is a factor of $2^{12}=4096$ in the worst case, which is not so prohibitively large as to make computation infeasible in realistic circumstances.

There is one final issue in that, as alluded to in \Cref{appendix:ft-condition}, in contrast to the preparation of Pauli eigenstates, magic states are usually stabilised by a set of Clifford operators that are not also Paulis. For example, the magic state $T \ket +$ is stabilised by the operator $e^{-\frac{i \pi}{4}}SX$. If the error combination $E'=e^{-\frac{i \pi}{4}}SX$ were to result in a circuit that purports to prepare this magic state $\ket T$, then the procedure we have so far outlined would incorrectly conclude that the error combination $E'$ is bad as the $X$ part of it is logical, whereas in reality the error combination $E'$ is good because as it stabilises the magic state it has no effect on it and is thus indistinguishable from no error having taken place.

We resolve this issue by considering the product of the error combination with all possible Cliffords $C_i$ that stabilise the magic state,
\begin{equation}
E''_i=E'C_i=E_s \ldots E_1 C_i,
\end{equation}
and we consider the error combination $E'$ to be good if $E''_i$ satisfies the procedure we have outlined for at least one choice of $i$.

In the case of preparing the usual magic state $T \ket +$ on a code with a single logical qubit, then the only Cliffords to consider are the identity and the operator $e^{-\frac{i \pi}{4}}SX$. However, as the number of logical qubits $k$ increases, there will be more Cliffords to consider, which may lead to an unfavourable scaling with $k$.

\section{Demonstration} \label{sec:demonstration}

\begin{table*}
\begin{tabular}{|c|c|c|c|c|c|c|c|}
\hline
 Name & Code  & No. of 2Q gates & No. of faults & $t_1$ & No. of combinations of two faults & $t_2$ & $t_f$ \\
\hline
 Steane & $\nkd{7}{1}{3}$    & 38 & 788 &  0.006 s& - & -      & 13 s\\
 $d=5$, unoptimised & $\nkd{17}{1}{5}$    &  298& 5040& 0.747 s& 12 662 112& 8m 3s& 11m 17s\\
 $d=5$, optimised & $\nkd{17}{1}{5}$    &  196& 3408 & 0.486 s& 5 781 600 & 3m 42s& 4m 55s\\
 $d=3$ & $\nkd{17}{1}{5}$    &  161& 2712 & 0.260 s& (100 292) & 4.05 s& 32.9 s\\
 Non-fault tolerant & $\nkd{17}{1}{5}$ &  24& (18)& 0.003 s&  -& -      & 12 s\\
\hline
\end{tabular}
\caption{Running times for Comb verifying the fault tolerance of circuits that prepare magic states $T \ket +$ to either distance three or distance five. The time $t_1$ is the time taken to check all single faults, $t_2$ is the time taken to check all combinations of two faults, and $t_f$ is the total wall time consumed by the program. The numbers in parentheses indicate the number of combinations of faults that were checked up to finding a bad combination. The time $t_2$ was omitted for the $d=3$ and the non-fault tolerant circuit as only single faults needed to be considered for these circuits. $t_f$ exceeds the sum of $t_1$ and $t_2$ owing to the time taken to perform other tasks such as propagating the faults through the circuits, which is not done in Comb but $t_f$ is included anyway to give an idea of general running times. The first two circuits prepare magic states in the Steane code and distance five colour code respectively using Chamberland's construction \cite{chamberlandFaulttolerantMagicState2019}. The third circuit, reproduced in \Cref{fig:circuit}, is obtained from the second by deleting flag ancillas from the circuit, and is hence much smaller, accounting for the reduced time taken to verify its fault tolerance. The time taken to generate the lookup tables for the codes in both cases was well under a second and does not contribute significantly to the total time $t_f$.}
\label{table:table}
\end{table*}

We use these theoretical insights to design and implement Comb \footnote{\url{https://github.com/Quantinuum/comb}}, an open-source library that iterates over a set of fault combinations arising from some CNOT + T circuit that purports to prepare a magic state fault-tolerantly and verifies that each of the fault combinations is good. It is designed to be fast so as to aid in the iterative design of fault tolerant magic state preparation circuits of distance greater than three. It takes as input a set of fault locations, a set of DCP operators that correspond to the output error that results from propagating each Pauli fault through the circuit, as well as a description of the magic state that the circuit prepares, that includes lookup tables to verify that the weight of any output error is small.

Comb is intended solely to perform the iteration over combinations of faults very quickly, and verify that they are either detectable or low weight. Consequently, in order to use it for a particular circuit, one must choose an appropriate set of fault locations and faults and propagate them through the circuit to obtain the DCP errors, since Comb itself cannot propagate faults through a circuit, and then supply them to Comb. One must also calculate the lookup tables using the algorithm outlined in \Cref{appendix:weight-lookup-tables}.

We demonstrate the performance of Comb in \Cref{table:table} when it comes to verifying the fault tolerance of several magic state preparation circuits. All the circuits tested have the general structure shown in \Cref{fig:chamberland-schematic}, based on the construction of~\cite{chamberlandFaulttolerantMagicState2019,chamberlandVeryLowOverhead2020a}.
For the $\nkd{17}{1}{5}$ code we consider several versions of the construction. One that is non-fault-tolerant and includes no logical stabiliser measurements or error detection cycles. A second that has circuit fault distance $d=3$ meaning it can handle a single fault but not two, this is achieved by appending one cycle of the logical stabiliser measurement and error detection. Finally, we consider two circuits that achieve the same circuit fault distance as the distance of the code $d=5$ by including two cycles of logical stabiliser measurements and the error detection.
A Dell Latitude 7430 laptop with a Intel(R) Core(TM) i7-1265U CPU and 16GB of RAM was used. However, as Comb is single-threaded at the time of writing the number of cores is not expected to have a bearing on the fraction of the running time spent in Comb, though the faults themselves were propagated in parallel to calculate the output errors.

For each circuit, the CNOT and T gates were assumed to be noisy two- and single-qubit gates respectively that are immediately followed by a fault location that may spawn faults equal to any Pauli on the support on the qubits involved in the gates. Comb was used to verify the fault tolerance criterion that each possible combination of $s$ faults (for appropriate $s\leq \lfloor \frac{d}{2} \rfloor$ chosen according to the distance of the code), leads to either a detectable error or an error of weight less than or equal to $s$ at the output.  

One important use case we envision for Comb is to help design fault-tolerant magic state preparation circuits with lower resource requirements, we give a simple illustrative example of that using the $\nkd{17}{1}{5}$ colour code. In \Cref{table:table} we show the performance of Comb on two $d=5$ variants of the magic state preparation circuit for the $\nkd{17}{1}{5}$ code. The smaller of these circuits was obtained from the first using a simple greedy iterative optimization process with Comb certifying at each step that the reduced circuit remained fault-tolerant. 
The final circuit, along with additional details of the optimisation, are given in \Cref{sec:appendix-17-1-5-optim}.

\begin{figure} 
\centering
\includegraphics[width=\columnwidth]{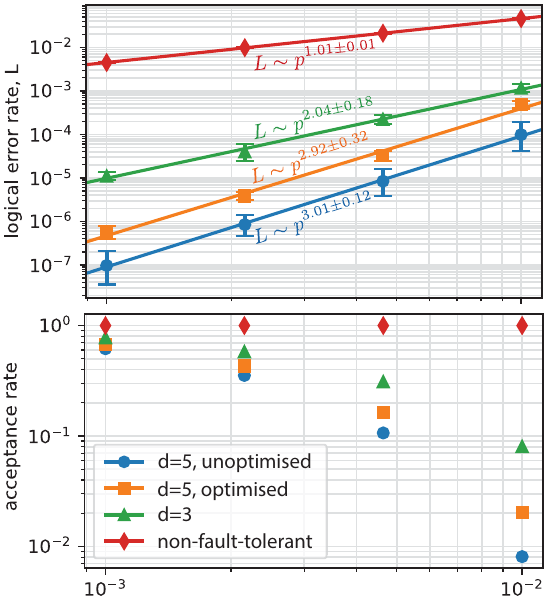}
\caption{
Numerically simulated logical error rates, $L$, and post-selection acceptance rates as a function of physical error rate $p$ for the magic state preparation circuits for the $\nkd{17}{1}{5}$ code tested by Comb in \Cref{table:table}.
In each case we see scaling consistent with Comb's findings: the non-fault-tolerant circuit has a logical error rate $L \sim p$, the $d=3$ variant has $L \sim p^2$ and the $d=5$ variants have $L \sim p^3$.
} \label{fig:numerics}
\end{figure}

As an independent check of Comb's correctness we perform noisy circuit simulations of all the $\nkd{17}{1}{5}$ circuits tested with Comb and compute their logical error rates, plotted as a function of physical error rate $p$ in \Cref{fig:numerics}. In each case we observe scaling of the logical error rates with the physical error rate that is consistent with the findings of Comb. We additionally plot the post-selection acceptance rates for each circuit.
These simulations were carried out using SymFT~\cite{fangSymFTUniversalFaultTolerant2026}. Additional details of the simulations are given in \Cref{sec:appendix-17-1-5-sims}.
It is interesting to note that in these simulations we see better logical error rate for the unoptimised distance 5 circuit than the optimised circuit. We attribute this to the simplicity of the error model used in the simulations, which does not include memory errors such as from coherent dephasing or qubit leakage. 
In experiments we would expect to see advantages from shorter depth resource state preparation circuits that are not captured in these simulations, including increased robustness to memory errors as well as decreased idling times for other logical qubits in the computation. 
The shorter circuit depths combined with improved acceptance rates has the potential to improve the rate of producing magic states while decreasing the number of ancilla qubits needed to manufacture them.

\section{Discussions and outlook} \label{sec:discussions-and-outlook}

In this paper, we make the case to explicitly restrict magic state preparation circuits to the CNOT + T gate set, and present Comb, a library that implements our algorithm insights and is designed to verify the fault tolerance of such circuits quickly.

One future direction is to use Comb to design new fault tolerant circuits, which is an area of active research. The distance-five magic state preparation circuit that is presented in \Cref{fig:circuit} achieves a reduction of over 30\% in the number of CNOT gates over the circuit it was derived from, and the fact that it was obtained simply by greedily deleting ancillas from another fault tolerant state preparation circuit gives us hope that there exist many fault tolerant magic state preparation circuits with much lower resource consumption than currently known distillation- or cultivation-based circuits, that may be verified using Comb.

Work is underway to analyse pytket \cite{Sivarajah_2021} circuits with Comb, which will greatly reduce the barrier to entry of using Comb for QEC researchers.

We anticipate that an autonomous agent, perhaps powered by AI, may be able to propose new magic state preparation circuits and then use Comb to check them. This suggests the possibility of closed-loop generation of magic state preparation circuits, which may lead to novel circuits that a human designer would struggle to propose in reasonable time.

There have been a plethora of recent results \cite{rodatzFloquetifyingStabiliserCodes2024a,poorUltraLowOverhead2025,rodatzFaultToleranceConstruction2026,ruschCompletenessFaultEquivalence2026} concerning fault tolerant rewrites for the ZX calculus, which allow one to change a fault-tolerant circuit using local rewrites that do not destroy its fault tolerance properties. Such techniques, relying as they do on local operations, are inherently scalable and promise to realise a practical way to reduce the resources that a fault tolerant quantum computation consumes at larger distances. To our knowledge, such results have been done exclusively on Clifford circuits to date. Our results concerning the practicality of the CNOT + T gate set for fault tolerant magic state preparation motivate the investigation of fault tolerant rewrites for ZX diagrams that correspond to CNOT + T circuits.

There exist many potential performance improvements to Comb that have not yet been implemented, such as use of parallelism, with both multiple cores and SIMD. We expect such improvements to have a great impact on performance, as verifying fault tolerance is inherently a parallel problem, as different combinations of faults can be divided into groups and checked independently in parallel.

Such performance improvements will not only aid in the exploration of distance-three and distance-five magic state preparation circuits, but will plausibly, combined with the use of HPC, make possible the exploration of distance-seven magic state preparation circuits, which is an exciting prospect.

\section{Acknowledgments}

The authors are employed by Quantinuum. We would like to thank Asmae Benhemou, Șerban Cercelescu, Ben Criger and Pablo Andrés-Martínez for productive discussions, Alec Edgington for invaluable help with C++, Marek Kochanowicz for useful insights regarding program optimisation, and Mateusz Kupper for assistance with the noisy simulations.

\bibliography{cnot-t-paper-references}

\appendix

\crefalias{section}{appendix}

\section{Counterexample to checking fault-tolerance by replacing $T$ gates with $S$ and $Z$} \label{sec:appendix-counterexample}

A trick that is sometimes used when designing and testing magic state preparation circuits is to replace the $T$ gates with $S$ or $Z$ gates \cite{gidneyCleanerMagicStates2023,gidneyMagicStateCultivation2024}. For some circuits (though not generically) this will replace the preparation of $T|+\rangle$ with $S|+\rangle$ or $Z|+\rangle$, without changing the circuit structure. Checking whether these CNOT + $S$ or CNOT + $Z$ circuits are fault-tolerant is much easier because they are Clifford circuits, though they are not preparing the desired magic state.

In this appendix we consider circuits $\mathcal{C}(Q)$ with the property that there are a set of $Q$ type gates in the circuit such that setting $Q=Z, \, S$ or $T$ prepares the logical state $Q |+\rangle$ and we ask what, if any, links we can draw between the fault-tolerance of $\mathcal{C}(S)$, $\mathcal{C}(Z)$ and $\mathcal{C}(T)$. An example of such a circuit is shown in Fig.~\ref{fig:counter-example-circ}(a), for the [[7,1,3]] Steane code.

\subsection{$\mathcal{C}(S)$ and $\mathcal{C}(Z)$ \underline{FT} $\not\Rightarrow \mathcal{C}(T)$ \underline{FT}}
We might hope that by finding a way to make $\mathcal{C}(S)$ and $\mathcal{C}(Z)$ fault-tolerant we will discover circuits that will also be fault-tolerant for $\mathcal{C}(T)$.

Here we present a circuit $\mathcal{C}(Q)$ where $\mathcal{C}(S)$ and $\mathcal{C}(Z)$ are fault-tolerant preparations of $S|+\rangle$ and $Z|+\rangle$ while $\mathcal{C}(T)$ is a non-fault-tolerant preparation of $T|+\rangle$.
This demonstrates that $\mathcal{C}(S)$ and $\mathcal{C}(Z)$ being fault-tolerant does not in general guarantee that $\mathcal{C}(T)$ is also.
This agrees with \cite{gidneyMagicStateCultivation2024}, which found in small statevector simulations that the logical error rate for preparing $T|+\rangle$ in a magic state cultivation scheme was about twice as high as $S|+\rangle$ or $Z|+\rangle$. 

\begin{figure*}[t]
\centering
\includegraphics[width=\textwidth]{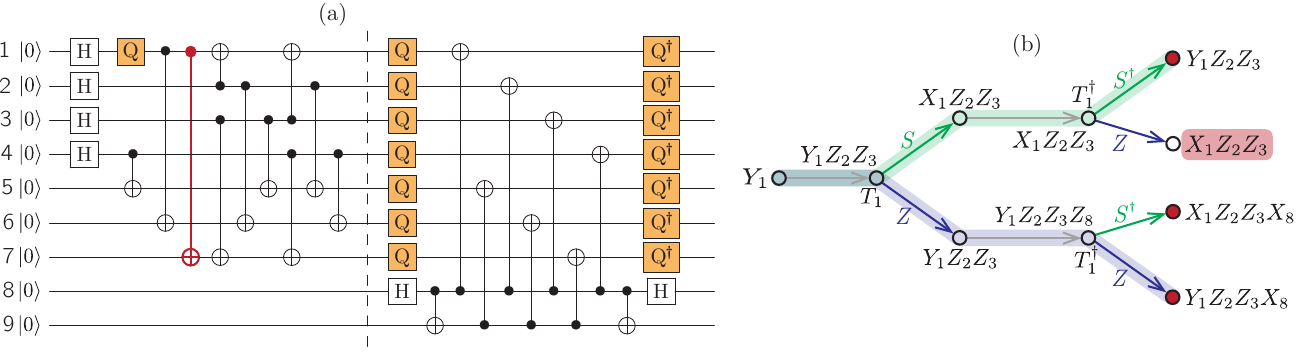}
\caption{\label{fig:app-counter-example-circ} (a) State preparation circuit for the Steane code $\mathcal{C}(Q)$ that prepares the logical $Q |+\rangle$ state when setting the highlighted gates to $Q=Z, \, S$ or $T$. The top 7 qubits are the code qubits and the bottom 2 qubits are ancillas. The ancillas are measured at the end and the circuit is post-selected on a +1 measurement outcome for both. The circuit is not fault-tolerant but can be made fault-tolerant for all $Q$ by measuring stabiliser elements at the end.
(b) Visualization of the Pauli propagation of a problematic fault for the $\mathcal{C}(T)$ circuit, which is missed by considering $\mathcal{C}(S)$ and $\mathcal{C}(Z)$. The error begins as a $YI$ error after the CNOT gate highlighted (bold, red) in (a). Clifford propagation is indicated with horizontal lines. Up to phases, which we neglect here, the $T$ gate tranforms the Pauli into a sum of two Paulis: one equivalent to the $T$ gate having been an $S$ and the other equivalent to it having been a $Z$. Here we indicate this by the Pauli branching into two branches. The $\mathcal{C}(S)$ and $\mathcal{C}(Z)$ circuit variants are equivalent to the top and bottom highlighted paths. If we modify the circuit in (a) to $\mathcal{C}'(Q)$ by measuring final stabilisers such that $\mathcal{C}'(S)$ and $\mathcal{C}'(Z)$ are fault-tolerant, problematic errors for $\mathcal{C}'(T)$ can arise on other branches of the propagation, such as the final $X_1 Z_2 Z_3$ error here on the $(S,Z)$ branch.
\label{fig:counter-example-circ}}
\end{figure*}

The circuit we consider is shown in Fig.~\ref{fig:counter-example-circ}(a), it is a non-fault-tolerant state preparation circuit for the [[7,1,3]] Steane code.
Setting the highlighted gates to $Q=Z, \, S$ or $T$ prepares the logical state $Q |+\rangle$.
It is composed of two parts: a non-fault tolerant encoding circuit, followed by a flagged fault-tolerant measurement of the logical stabiliser~\cite{chamberlandFaulttolerantMagicState2019}. The logical stabiliser measurement catches logical errors. Nevertheless, it is non-fault-tolerant because there are single faults that propagate to errors on the output with weight 2 or higher.

We can identify minimal size sets of stabilisers that make the circuits fault-tolerant if we measure them at the end of the circuits. The $Z |+\rangle$ circuit can be made fault-tolerant by measuring a single stabilier element:
\begin{gather*}
IXXIIXX .
\end{gather*} 
For the $S |+\rangle$ circuit there are no single stabilisers that catch all bad errors, but there are pairs of stabilisers that will make the circuit fault-tolerant:
\begin{gather*}
YYIYIIY , \, YIYYIYI \\
YYIYIIY , \, IYYIIYY \\
ZZIZIIZ , \, IYYIIYY \\
YIYYIYI , \, IYYIIYY \\
ZIZZIZI , \, IYYIIYY \\
IYYIIYY , \, YYIIYYI \\
IYYIIYY , \, ZZIIZZI \\
IYYIIYY , \, YIYIYIY \\
IYYIIYY , \, ZIZIZIZ \\
YYIIYYI , \, YIYIYIY \, .
\end{gather*}
Finally, there are triples of stabilisers that can be appended to the $T |+\rangle$ circuit to make it fault-tolerant:
\begin{gather*}
ZZIZIIZ , \, IIIZZZZ , \, IXXIIXX \\
ZZIZIIZ , \, IYYIIYY , \, XXIIXXI \\
ZZIZIIZ , \, ZZIIZZI , \, IXXIIXX \\
IIIZZZZ , \, ZZIIZZI , \, IXXIIXX \, .
\end{gather*}
These stabiliser elements can be measured fault-tolerantly at the end of the circuits using flag gadgets~\cite{reichardtFaulttolerantQuantumError2021}.

Appending three fault-tolerant measurement gadgets to the end of the circuits to measure the $IXXIIXX$ , $YYIYIIY$ \& $YIYYIYI$ stabilisers yields a circuit $\mathcal{C}(Q)$ such that $\mathcal{C}(S)$ and $\mathcal{C}(Z)$ are fault-tolerant but $\mathcal{C}(T)$ is not. 

Although there are triples of stabilisers that we can measure to make the circuit in Fig.~\ref{fig:counter-example-circ}(a) a fault-tolerant preparation of $T|+\rangle$ this choice made by considering the $S$ and $Z$ variants is not sufficient.

An intuitive understanding of this failure can be gained by considering the Pauli propagation of a problematic fault, Fig.~\ref{fig:counter-example-circ}(b). A $YI$ error after the CNOT gate highlighted (bold, red) in Fig.~\ref{fig:counter-example-circ}(a) propagates through the subsequent CNOT gates to become $YZZIIIIII$ before the first layer of $T$ gates. The $T$ gate on qubit 1 acts as a linear combination of an $S$ gate and a $Z$ gate, which we draw as two paths leading to $XZZIIIIII$ and $YZZIIIIII$.
To see this, consider the conjugation of a Pauli $P$ that anticommutes with $Z$ through a $T$ gate
\begin{equation*}
    T P T^\dagger = S P \propto (1 - iZ)P = - ZPZ -i SPS^\dagger .
\end{equation*}
For our purposes here we can neglect the phases on these terms.
Through the CNOTs of the logical stabiliser measurement the second term gains a $Z$ on the first ancilla to become $YZZIIIIZI$, this term is a product of a $Z$ logical and a single qubit $X$ error and so the logical stabiliser measurement has been triggered. This term splits again at the second layer of $T^\dagger$ gates but this does not remove the support on the ancilla and so both are detected when we measure the ancillas. The $XZZIIIIII$ term, in contrast, does not trigger the logical stabiliser measurement and splits at the $T^\dagger$ layer into two terms $YZZIIIIII$ and $XZZIIIIII$. The first of these anti-commutes with our choice of final stabiliser measurements and is detected, but the second commutes with the stabilisers and so is undetected at the output of the circuit, becoming a high weight bad error. By inspecting what paths have given rise to detected and undetected errors it is clear what has happened: the $(S, S^\dagger)$ branch is detected, we have guaranteed this when considering $\mathcal{C}(S)$. The problematic error has arisen from the $(S, Z)$ branch, which is absent when considering $\mathcal{C}(S)$ and $\mathcal{C}(Z)$.

\subsection{Does $\mathcal{C}(S)$ and $\mathcal{C}(Z)$ \underline{non-FT} $\Rightarrow$ $\mathcal{C}(T)$ \underline{non-FT}?}

If $\mathcal{C}(S)$ and $\mathcal{C}(Z)$ being non-fault-tolerant guaranteed that $\mathcal{C}(T)$ also cannot be fault-tolerant this would still be a useful property as it would offer a fast rejection method for candidate circuits. Using the fact that $\mathcal{C}(S)$ and $\mathcal{C}(Z)$ are Clifford circuits, their fault-tolerance can be checked with standard Pauli propagation and weight minimisation methods. We could quickly test $\mathcal{C}(S)$ and $\mathcal{C}(Z)$, and only proceed to the more expensive testing of $\mathcal{C}(T)$ if they both pass.

We leave this as an open problem. We are not able to provide a counter example, but we cannot prove that it must always be true.

\section{Fault tolerance condition} \label{appendix:ft-condition}

Here we show that the condition given for the fault-tolerance of CNOT + T state preparation circuits in section~\ref{sec:verifying-fault-tolerance:defn} is equivalent to those commonly given in the literature~\cite{chamberlandFaulttolerantMagicState2019,gottesmanSurvivingQuantumComputer2024}. We specifically compare to the definition given by Gottesman in~\cite{gottesmanSurvivingQuantumComputer2024}.

For completeness, the condition we use is: a CNOT + T state preparation circuit is fault-tolerant to distance $d$ if it obeys:
\begin{enumerate}
    \item[($\star$)] any combination of $s  \leq t=\floor{\frac{d}{2}}$ faults produces an output error that is either detected by a pattern of unexpected ancilla measurements or is a sum of Pauli strings each of weight at most $s$.
\end{enumerate}
The weight of a Pauli string must be taken modulo the stabilisers. This is achieved with a weight lookup table, as discussed in appendix~\ref{appendix:weight-lookup-tables}.

The fault-tolerance conditions for state preparation circuits presented in~\cite{gottesmanSurvivingQuantumComputer2024} make use of \emph{r-filters} and \emph{ideal decoders}. This machinery is more general than we needed so we avoided it in our definition, opting instead for a simpler (but equivalent) condition based on Pauli weights.

An \emph{r-filter} for a QEC code projects onto the subspace of errors spanned by states $E |\psi\rangle$ where $|\psi\rangle$ is any codeword and $E$ is a Pauli error of weight at most $r$. We use the terminology \emph{passing an r-filter} to mean that the state is unchanged by applying an $r$-filter.

An \emph{ideal decoder} is a quantum channel that takes an encoded state, noiselessly extracts the syndrome of any errors before correcting them and returning the decoded logical state. If the errors on the encoded state are uncorrectable the ideal decoder will return a state with a logical error.

Fault-tolerance for a state-preparation circuit is defined in \cite{gottesmanSurvivingQuantumComputer2024} by the two properties:
\begin{enumerate}
    \item[(1)] \emph{Preparation Error Propagation:} any combination of $s \leq t$ faults passes an $s$-filter,
    \item[(2)] \emph{Preparation Correctness:} the noisy output state will be decoded by an ideal decoder to the logical state that the noiseless circuit would prepare, provided $s  \leq t$ faults occur,
\end{enumerate}
where $t=\floor{\frac{d}{2}}$ is the number of correctable errors. 

We first show that if ($\star$) is satisfied then (1) is satisfied. 
Let us write the output DCP error as
\begin{equation}
E = \sum_i c_i P_i \, ,
\end{equation}
where the $P_i$ are Pauli strings.
The weight $E$ is taken modulo the stabilisers. To be explicit the weight is minimised with respect to the full stabiliser group of the code $\mathcal{S}$ enlarged with the logical Clifford $C$ that the noiseless output state is the +1 eigenstate of, e.g. for the $T|+\rangle$ logical state $C = \overline{T} \, \overline{X} \, \overline{T}^\dagger = \overline{S} \, \overline{X}$. 
Consider the minimisation under $\mathcal{S}$ first, each $P_i$ can be written
\begin{equation}
P_i = Q_i \, S_i
\end{equation}
where $Q_i$ is the Pauli string with minimum weight and $S_i$ is an element of the stabiliser group.
The action of $E$ on the noiseless encoded output state $|\overline{\Psi}\rangle$ is
\begin{align}
E |\overline{\Psi}\rangle &= \sum_i c_i P_i |\overline{\Psi}\rangle \nonumber \\
&= \sum_i c_i Q_i S_i |\overline{\Psi}\rangle \nonumber \\
&= \sum_i c_i Q_i |\overline{\Psi}\rangle \, , \label{eqn:appFT-output-state}
\end{align}
where we have used the fact that $|\overline{\Psi}\rangle$ is stabilised by $S_i$. 
The noiseless state $|\overline{\Psi}\rangle$ is also stabilised by $C$ so we can consider the alternative DCP error 
\begin{equation}
E' = E \, C = \sum_j d_j P'_j \, ,
\end{equation}
where the sum will generically contain a different number of Paulis. Minimising the weight of each $P'_j$ under the stabilisers produces an alternative set of minimum weight Paulis $\{ Q'_j \}$.
Condition ($\star$) requires that the weight of all the Paulis $\{Q_i\}$, or all the $\{ Q'_j \}$, is at most $s$ when $s \leq t$ faults occur. If this condition holds then the output state Eqn.~\eqref{eqn:appFT-output-state} will pass an $s$-filter, since $|\overline{\Psi}\rangle$ is a codeword, hence condition (1) is satisfied.

Next we show that if ($\star$) is satisfied then (2) is satisfied. 
If the weight of all the $Q_i$ is at most $s \leq t$ then all the $Q_i$ are correctable errors \cite{gottesmanSurvivingQuantumComputer2024}. In this case an ideal decoder is sure to decode the output state Eqn.~\eqref{eqn:appFT-output-state} to $|\Psi\rangle$, the logical state of the noiseless circuit, and so (2) is satisfied.

\section{Diagonal coherent error analysis} \label{sec:diagonal-coherent-error-analysis}
A great advantage of DCP errors is that they are able to compactly describe coherent errors which, when expressed as a sum of Paulis, may be very large.

In fault tolerant analysis, one is interested in whether errors are detected by ancilla measurements, and if not, what the residual conditional error on the remaining qubits is. It is known how to do this efficiently for errors that are Pauli. For a general coherent error, one is compelled to expand the coherent error as a sum of Paulis and check each term in the sum in turn. If the sum is very large, then even initially writing down the sum (or storing it memory on a computer) for a single coherent error will be infeasible, let alone considering a very large number of errors.

In this section, we will show that given a coherent error prior to any measurements that is DCP, it is possible to efficiently calculate and describe the conditional error that results from postselecting on the results of ancilla measurements, without having to expand any sums explicitly. This result is of great importance in ensuring that DCP errors can be verified to be good or bad in a reasonable timeframe.

\subsection{Operator state duality}
As it is possible to deal with the Pauli X part of a DCP error in much the same way as a Pauli error, we assume for the rest of this section, without loss of generality that the error $E$ that results immediately prior to any ancilla measurements is strictly diagonal Clifford.

As discussed, we are interested in dealing with the error $E$ without having to express it directly as a sum of Paulis. In order to do so, we will make use of a operator-state duality. We define the {\it error state} $\ket{\psi_E}$ to be the state
\begin{equation}
\ket{\psi_E} = E \ket{+}^{\otimes n}.
\end{equation}

The error state can be defined not only for the error $E$ that is diagonal Clifford, but for all diagonal operators. The error state contains all information about the error $E$; it is possible to reconstruct the operator $E$ from the state $\ket{\psi_E}$ by observing that the coefficients of the state $\ket{\psi_E}$ in the computational basis form the diagonal of the operator $E$ in that basis.

It will turn out that many operations of interest that one would like to carry out on the coherent error $E$, such as determining the conditional coherent error $E'$ that results from postselecting on an ancilla measurement, are best expressed in terms of operations on the corresponding error state $\ket{\psi_E}$.

Before turning to them, we will first note that since the error $E$ is diagonal Clifford, the error state $\ket{\psi_E}$ is a Pauli stabiliser state, as it is defined as the application of a Clifford $E$ to the state $\ket{+}^{\otimes n}$, which is a Pauli stabiliser state. Furthermore, a choice of stabiliser generators for the error state $\ket{\psi_E}$ can be determined easily from a description of the diagonal Clifford $E$, as they are simply the images of the Pauli operators $X_i$ under conjugation by the error $E$.

\subsection{Postselection}
In order to detect such diagonal coherent errors, one or more ancillas may be measured in the $X$ basis, and the state discarded if one of the measurement results, or a parity of them, is non-trivial. We denote $P_X$ to be the Pauli of $X$ type corresponding to one such measurement.

Writing the error $E$ as a sum of diagonal Paulis,
\begin{equation} \label{eq:e-sum-1}
E = \sum_{i=1}^N c_i P_i,
\end{equation}
the error $E'$ that results will be of the form
\begin{equation}  \label{eq:e-sum-2}
E' = \sum_{i=1| [ P_i, P_X ] =0}^N c_i P_i,
\end{equation}
which is to say that the errors that anticommute with the measurement operator are detected, while those that do not are undetected and survive to form the coherent error conditional on postselection. 

Eliminating terms from $E$ in this drastic manner means that $E'$ may not necessarily be Clifford; indeed it may vanish altogether. Despite this, it is certainly diagonal, which means that it possesses an error state $\ket{\psi_{E'}}$, and there exists a simple relation between the error states $\ket{\psi_{E}}$ and $\ket{\psi_{E'}}$.

Applying \Cref{eq:e-sum-1} and \Cref{eq:e-sum-2} to the state $\ket{+}^{\otimes n}$ gives that
\begin{equation} \label{eq:e-sum-3}
\ket{\psi_E} = \sum_{i=1}^N c_i P_i\ket{+}^{\otimes n}, 
\end{equation}
and
\begin{equation} \label{eq:e-sum-4}
\ket{\psi_{E'}} = \sum_{i=1| [ P_i, P_X ] =0}^N c_i P_i\ket{+}^{\otimes n}.
\end{equation}

The terms $P_i \ket{+}^{\otimes n}$ that appear in \Cref{eq:e-sum-4} are those terms appearing in \Cref{eq:e-sum-3} for which $P_i$ commutes with $P_X$. These terms are precisely the positive eigenstates of $P_X$, which means that they are preserved under the projector onto the positive eigenspace of $P_X$, while the unwanted terms are eliminated by this projector.

Concretely, we get that
\begin{equation}
\ket{\psi_{E'}} = \frac{I + P_X}{2} \ket{\psi_E}.
\end{equation}

As previously discussed, the error state $\ket{\psi_E}$ is a Pauli stabiliser state. The above equation gives that the state $\ket{\psi_{E'}}$ is equal to that which results from measuring the Pauli operator $P_X$ to be +1 on the state $\ket{\psi_E}$. But as the state $\ket{\psi_E}$ is a Pauli stabiliser state, a choice of stabiliser generators of the state $\ket{\psi_E'}$ can be evaluated efficiently by updating those of the state $\ket{\psi_E}$\cite{gottesmanSurvivingQuantumComputer2024}. We note also that the resulting state may vanish, in which case the conditional error $E'$ vanishes also. This case corresponds to the situation in which the error $E$ is completely detected by the measurement.

Although the error $E'$ is not necessarily Clifford, its error state $\ket{\psi_{E'}}$ is a Pauli stabiliser state just as the state $\ket{\psi_E}$ is. This means that we can apply the above technique again to $\ket{\psi_{E'}}$ to evaluate the error state that results from postselecting on any number of ancilla measurements, not just one, simply by measuring the corresponding measurement operators in turn on the error state.

In this way, one can describe in terms of its error state the conditional coherent error $E_C$ that results from all postselections made in the X basis. Such a state will necessarily have much fewer terms in its expansion than the original error $E$.

\subsection{Canonicalisation}
The conditional coherent error $E_C$, when expressed as a linear combination of diagonal Paulis, may contain distinct terms which are equal up to a stabiliser. It is important to eliminate such terms in the sum by stabiliser multiplication and cancellation in order to accurately evaluate whether the error is good (low weight) or not.

To illustrate this more clearly, we give a hypothetical example. A circuit that purports to prepare the magic state $T \ket +$ in the Steane code might have a coherent error, after postselection, of the form
\begin{equation}
E_C = ZZZIIII - ZZZZZZZ.
\end{equation}

Both of these terms are logical operators, and so they cannot be equivalent to a weight one error up to multiplication by any $Z$ stabiliser, since the code is distance three. Consequently, they are both undetectable bad errors and a circuit that produced either of these errors in the presence of a single fault would have to be rejected as not fault tolerant.

However, they are both equal after multiplication by the stabiliser generator $IIIZZZZ$. Consequently, the error $E_C$ vanishes, which necessarily means that it is good, as a vanishing error is one that was already detected by postselecting on an ancilla measurement.

Consequently, when considering coherent errors, one must take care to ensure that errors that are equivalent up to multiplication by a stabiliser have already been cancelled out in the sum, otherwise checking each term in the sum in turn would lead to a spurious conclusion that some good coherent errors are bad.

Surprisingly, it is possible to perform such a cancellation by measuring the error state $\ket{\psi_{E_C}}$ without having to expand the sum at all.

We note first that, following \Cref{appendix:weight-lookup-tables}, every Pauli operator has a unique canonical form that results from multiplying by a choice of stabiliser generators. Writing the operator
\begin{equation}
E_C = \sum_{i=1}^N c_i P_i,
\end{equation}
we note that if every operator $P_i$ is replaced by its canonical form, then all cancellations in the sum will take place automatically, as operators that are equivalent up to a stabiliser will be mapped to the same canonical form.

Every term in the sum $E_C$ can be canonicalised by multiplying on the left by $E_C$ by an appropriate choice of operators. In order to help illustrate this, we will give a worked example. In order to canonicalise operators with respect to the Steane code, operators which have support on the first qubit must be multiplied by the stabiliser generator $S_Z=ZIZIZIZ$, and those that do not are left unchanged. Doing the same for subsequent stabiliser generators of the Steane code ensures that all operators have vanishing support on the first, second and fourth qubits, and are so left in canonical form.

Considering for now only the effect of canonicalising on the first qubit, we must separate the error $E_C$ into a part $E_C^+$ which has trivial support on the first qubit , and a part $E_C^-$ that doesn't. It will prove simpler to discuss these operators in terms of their error states. The error states according to these operators can be obtained straightforwardly by projecting into the positive and negative eigenspaces of $X_1$ respectively,
\begin{align}
\ket{\psi_{E^+_C}} = \frac{I + X_1}{2} \ket{\psi_{E_C}}, \label{eq:plus-e}\\
\ket{\psi_{E^-_C}} = \frac{I - X_1}{2} \ket{\psi_{E_C}}. \label{eq:minus-e}
\end{align}
The terms in $E_C^+$ have vanishing support on the first qubit and do not need to be changed. However, all the terms in $E_C^-$ have support on the first qubit, and must all be multiplied by $S_Z$ in order to canonicalise them. This can be accomplished by applying the operator $S_Z$ to the error state $\ket{\psi_{E_C^-}}$. Once this is done, both halves of the sum can be recombined. The error $E'_C$ that results is
\begin{equation}
\ket{\psi_{E'_C}} = \ket{\psi_{E^+_C}} + S_Z \ket{\psi_{E^-_C}}.
\end{equation}

Substituting \Cref{eq:plus-e} and \Cref{eq:minus-e} gives that

\begin{equation}
 \ket{\psi_{E'_C}} = \left( \frac{I + X_1}{2} + S_Z\frac{I - X_1}{2}\right) \ket{\psi_{E_C}}.
\end{equation}

This expression can be simplified by observing that $S_Z$ anticommutes with $X_1$, as $S_Z$ has support on the first qubit. One gets, after pushing $S_Z$ past $X_1$ and factorising, that
\begin{equation}
 \ket{\psi_{E'_C}} = \frac{I + X_1}{2} \frac{I + S_Z }{2} \ket{\psi_{E_C}}.
 \end{equation}

This corresponds to measuring two Pauli operators to be +1 on the error state $\ket{\psi_{E_C}}$, and generalises to canonicalising on any qubit in any CSS code, not just the Steane code.

As the state $\ket{\psi_{E_C}}$ is a Pauli stabiliser state, just as before the state $\ket{\psi_{E'_C}}$ that results from measuring Paulis on it is also a Pauli stabiliser state, and a description of it can be obtained efficiently.
\subsection{Error states to cosets}
In this way, one can efficiently determine the final coherent error $E_F$ that results from postselecting on all relevant ancilla measurements and eliminating terms that are equal up to a stabiliser multiplication. The next step is to expand the error $E_F$ as a linear combination of Paulis and ensure that each of them is low weight.

The previous results have shown how to efficiently obtain a description of the error $E_F$ in terms of its error state $\ket {\psi_{E_F}}$. We now turn to the problem of obtaining an expansion of $E_F$ in terms of Paulis from its error state. Although the number of terms in the sum will turn out to be exponential in some parameter, they all belong to the coset of a vector space which can be described and calculated efficiently.

Let us first relate the expansion of $E_F$ to the error state. Writing the error
\begin{equation}
E_F = \sum_{i=1}^N c_i P_i.
\end{equation}
as a linear combination of distinct Paulis $P_i$ of Z type, and applying both sides to the state $\ket{+}^{\otimes n}$, we get that
\begin{equation}
\ket{\psi_{E_F}} = \sum_{i=1}^N c_i P_i \ket{+}^{\otimes n}.
\end{equation}

Consider the states $P_i \ket{+}^{\otimes n}$. The effect of $P_i$ on the state $\ket{+}^{\otimes n}$ is to flip $\ket +$ to $\ket -$ on qubits on which $P_i$ has support, and leave it invariant otherwise. Consequently, one can read off $P_i$ from the patterns of $+$ and $-$ on such states. This means that all information about the sum - in particular the Paulis $P_i$ that make up the sum - can be obtained by expressing the error state $\ket{\psi_{E_F}}$ in the basis of X stabiliser states $B=\{ \ket{+}, \ket{-} \}^{\otimes n}$ and seeing which basis terms form the combination.

Since the error state $\ket{\psi_{E_F}}$ is a Pauli stabiliser state, it is possible to compactly characterise the set of diagonal Paulis $P_i$.

We first fix a choice of stabiliser generators $S_1, ..., S_n$ for the state $\ket{\psi_{E_F}}$.
In order to make contact with the basis $B$, we will fix some X basis state $\ket{C} \in B$ and write

\begin{equation} \label{eq:projector}
\ket{\psi_{E_F}} \propto \prod_{i=1}^n \frac{I + S_i}{2} \ket{C},
\end{equation}
where we have related the error state $\ket{\psi_{E_F}}$ to an X computational basis state $\ket{C}$ by applying the projector onto the error state $\ket{\psi_{E_F}}$ to the state $\ket C$. As the X computational basis states are a basis, there must always exist at least one choice of state $\ket C$ such that the constant of proportionality is not zero i.e. at least one state $\ket{C}$ must have non-zero overlap with the error state $\ket{\psi_{E_F}}$. We will turn later to the question of how to choose the state $\ket C$, and assume for now that it has been chosen appropriately by some means.

Let us now consider the effect of projecting the state $\ket{C}$ onto the positive eigenspace of the stabiliser generators $S_i$. In order to simplify the analysis, we will consider two special cases. If $S_i$ is of Z type and not the identity, then one gets that
\begin{equation}
\frac{I + S_i}{2} \ket{C} = \frac{\ket{C} + S_i\ket{ C}}{2},
\end{equation}
where the state $S_i \ket{C}$ is a X computational basis state that is distinct from $\ket{C}$.

If $S_i$ is of X type, then $S_i \ket{C}$ will be equal to $\ket{C}$ up to a sign. Correspondingly, acting on $\ket C$ with the projector $\frac{I + S_i}{2}$ will either give zero or the state $\ket{C}$, depending on the sign.

Of course, the stabiliser generators $S_i$ are not known to be of either X or Z type, and are unlikely to be. However, armed with these observations, we can cast the stabiliser generators $S_i$ into a different set $S'_i$ via Gaussian elimination, that will allow us to apply these observations.

Concretely, we perform Gaussian elimination on the Z parts of the stabiliser generators $S_i$. A new set of stabiliser generators $S'_i$ stabilising the same state is produced, such that for some known $r$, for all $i \leq r$ the Z parts of the stabiliser generators $S'_i$ are linearly independent, and for all $i >r$, the Z parts of the stabiliser generators $S'_i$ vanish.

Substituting the new set of stabiliser generators $S'_i$ into \Cref{eq:projector}, we get
\begin{equation}
\ket{\psi_{E_F}} \propto \left(\prod_{i=1}^r \frac{I + S'_i}{2} \right) \left(\prod_{i=r+1}^n \frac{I + S'_i}{2} \right) \ket{C}.
\end{equation}

First, we apply the projectors corresponding to $i > r$ to the state $\ket C$. The stabiliser generators $S'_i$ are all of X type by construction, and so as discussed they either eliminate the state $\ket{C}$ or preserve it. It is always possible to choose the state $\ket C$ such that none of the projections eliminate the state by writing $\ket C = P_C \ket{+}^{\otimes n}$, where $P_C$ is some Pauli of Z type, and choosing $P_C$ to anticommute or commute with $S'_i$ as appropriate to ensure that the projections do not eliminate the state $\ket C$. This can always be done efficiently by solving a linear system of equations over $\mathbb F_2$.

Once this has been done, the projections act trivially on the state $\ket C$ and we are left with
\begin{align}
\ket{\psi_{E_F}} &\propto \left(\prod_{i=1}^r \frac{I + S'_i}{2} \right) \ket{C} \\
&= \frac{1}{2^r} \sum_{a_1,...a_r \in \{0,1\}} S'_1{^{a_1}} ... S'_r{^{a_r}} \ket{C}.
\end{align}

The operators $S'_1{^{a_1}} ... S'_n{^{a_r}}$ that appear in the sum are not necessarily of either X or Z type, so when they act on the state $\ket C$ they can both flip the sign and take $\ket C$ to a different X computational basis state $\ket{C_{a_1, ...a_r}}$, depending on the Z parts of the operator $S'_1{^{a_1}} ... S'_r{^{a_r}}$. However, as the Z parts of the operators $S'_i$ are all linearly independent by construction, the states $\ket{C_{a_1, ...a_n}}$ are all unique, meaning that there can be no cancellations in the resulting sum. Furthermore, the states $\ket{C_{a_1, ...a_r}}$ can be obtained by acting on the state $\ket C = P_C \ket{+}^{\otimes n}$ by all the $2^r$ ways of forming products of the Z parts of the stabiliser generators $S'_i$.

Consequently, one gets that the coherent error $E_F$ can be expressed as a linear combination of $2^r$ Paulis that form the coset of a vector space
\begin{equation}
S = P_C\langle S'^{Z}_1, ..., S'^{Z}_r \rangle,
\end{equation}
where the operator $S'^{Z}_i$ is the Z part of the stabiliser generator $S'_i$.

We conclude by remarking that although the number of errors in the coherent sum is $2^r$, it is at least possible to calculate and describe the set efficiently, and that the elements of the coset may be iterated over in constant memory via a Gray code for example.
\section{Weight lookup tables} \label{appendix:weight-lookup-tables}
We make liberal use of lookup tables to fetch the weight of an operator up to multiplications by a stabiliser from some code. In this Appendix, we discuss novel theoretical and technical work related to lookup tables that is key to the performance of Comb and may be of general interest to those working with stabiliser codes.

For simplicity of discussion, we will limit ourselves to CSS codes as we do not consider non-CSS codes elsewhere in this work, though much of the discussion does generalise to non-CSS codes.
\subsection{Lookup tables}
A lookup table is a simple data structure that stores the result of some computation, so that it need not be computed more than once. Given a computable but expensive function $f$ that acts on a finite set $R$ of size $N$, we produce a lookup table for $f$ by first labelling the members $ r_i \in R: 0 \leq i < N$ and then computing and storing the values $f(r_i)$ in an array of size $N$. Subsequent evaluations of $f$ may be made by looking up the appropriate entry in the array, which is an operation that takes constant time.

This dramatic saving in time naturally comes at the cost of a great increase in space. If the set $R$ is too large, it may be too impractical to store it in memory. Furthermore, we must evaluate $f$ on every element of $R$ in order to form the lookup table. This will not be worthwhile unless the number of evaluations of $f$ is likely to exceed $N$. Since by assumption $f$ is expensive to evaluate on individual elements, then evaluating $f$ on the entire set $R$ is likely to be very expensive indeed. However, if $R$ is known, then the lookup table only needs to be computed once, and depending on the details of evaluating $f$, forming the lookup table may not be prohibitively expensive.

We argue that lookup tables are appropriate to compute the weight of operators modulo a stabiliser code of reasonable size. Firstly, determining the weight of an operator $P$ modulo a stabiliser code is a computationally costly problem. Given a code with $s$ stabiliser generators of a code, one way to compute the weight of $P$ is to form all possible products of $P$ with a stabiliser and find the minimum weight over the products. However, as there are $2^s$ different stabilisers, this is not very efficient, and prohibitive for codes of even moderate size. We note also that, for the purposes of checking fault tolerance, the number of output error combinations that must be considered is extremely large, and accordingly we expect to check the weight of a very large number of operators. Any potential saving in the cost of evaluating the weight of an operator must therefore be considered.

We present two results that greatly facilitate the use of tables to look up the weight of an operator; firstly, a trick to compress the size of a lookup table, and secondly an algorithm to compute lookup tables in reasonable time.

\subsection{Compression trick} \label{sec:compression-trick}
Given a $\nkd{n}{k}{d}$ CSS code, we define the CSS weight lookup tables $T^X_i$ and $T^Z_i$,
\begin{align}
T^X_i(P_X) &= 1 \text{ if } \min \left\{ h(P_XS_X):S_X \in C_X\right\} \leq i \\ \nonumber
&= 0 \text{ otherwise},\\ \nonumber
T^Z_i(P_Z) &= 1 \text{ if } \min \left\{ h(P_ZS_Z):S_Z \in C_Z\right\} \leq i , \\ \nonumber
&= 0 \text{ otherwise},\\ \nonumber
\end{align}
where $P_X$ and $P_Z$ are Paulis of $X$ and $Z$ type respectively, $h$ is the Hamming weight, $C_X$ and $C_Z$ are the spaces spanned by the $X$ and $Z$ stabilisers of the code $C$ respectively.
One can verify whether the weight of a Pauli $P_X$ of $X$ type is of weight lower than or equal to $s$ by looking up the appropriate entry in the table $T^X_i(P_X)$, and similarly for a Pauli $P_Z$ of $Z$ type.

In practice, we will limit ourselves to the case $i \leq \lfloor \frac{d}{2} \rfloor$, since this is sufficient for verifying fault tolerance.

We define the weight lookup tables this way for two reasons: first of all, when we check whether a given output error combination is good, we are not interested so much in the weight of an output error combination as much as verifying that its weight is below some threshold $s$. Consequently, the weight lookup tables will contain all the information - one bit per Pauli - that is required and no more, which is optimal from the perspective of storing information.

The Paulis $P_X$ and $P_Z$ can be put into correspondence with bit strings of length $n$, which mean that each of these lookup tables are of size $2^n$. Given a code of reasonable size such as the Golay code, with $n=23$ and $d=7$, each of the six lookup tables will occupy 1MiB of space. These can easily fit into memory, which illustrates the practicality of lookup tables for this problem.

However, one can compress the lookup tables by a considerable degree by casting the stabiliser generators of a code into a particular form.

To use the Golay code as an example, its stabiliser generators of $Z$ type can be written,
\begin{verbatim}
ZIZIIZIIZZZZZIIIIIIIIII
ZZZZIZZIZIIIIZIIIIIIIII
IZZZZIZZIZIIIIZIIIIIIII
IIZZZZIZZIZIIIIZIIIIIII
IIIZZZZIZZIZIIIIZIIIIII
ZIZIZIZZZIIZIIIIIZIIIII
ZZZZIIIZIIZZIIIIIIZIIII
ZZIZZZIIIZZIIIIIIIIZIII
IZZIZZZIIIZZIIIIIIIIZII
ZIIZIIZZZZZIIIIIIIIIIZI
IZIIZIIZZZZZIIIIIIIIIIZ
\end{verbatim}
where the tableau restricted to the final eleven qubits is the identity. We note that this can be done for any stabiliser code up to a qubit permutation by performing Gaussian elimination.

Given an operator $P_Z$ on 23 qubits, one can multiply $P_Z$ by these stabiliser generators as appropriate to eliminate any $Z$ operators on the last eleven qubits, thus obtaining another operator $P'_Z$ which acts non-trivially on at most the first twelve qubits. We will refer to $P'_Z$ as the {\it canonical form} of $P_Z$.

Since multiplying an operator by a stabiliser preserves its weight modulo the stabiliser code, instead of using $P_Z$ to index into the lookup table we can instead use $P'_Z$. This means that the lookup tables now each occupy $2^{12}$ bits, or 512 bytes, which is a massive saving, and makes the use of lookup tables practical for larger codes.

This saving in space is in fact likely to deliver a saving in time. A lookup table of 512 bytes can fit into the L1 cache of many CPUs, which, assuming that the rest of the program does not occupy too much space, is likely to make memory accesses considerably faster.
\subsection{Computing a weight lookup table}
The compression trick discussed in the last section gives rise to an inductive algorithm for computing the lookup tables $T^Z_i(P_Z)$. Computing the tables $T^X_i(P_X)$ will follow by symmetry.

Considering first the base case $i=0$, we observe that $T^Z_0(P'_Z)$ is 1 if and only if $P'_Z$ is the identity. This is because only stabilisers may have weight zero modulo the code, and all stabilisers have the identity as their canonical form.

We will proceed to generate the tables $T^Z_i(P'_Z)$ inductively. Given tables $T^Z_0,\ldots T^Z_i$, we will show how to generate table $T^Z_{i+1}$.

Iterate over all elements $P$ such that $T^Z_{i}(P)=1$ {\it i.e.} those that have weight less than or equal to $i$ modulo the code. For each $1 \leq j \leq n$, we form the product $P Z_j$. By construction, $P Z_j$ must have weight less than or equal to $i+1$. We then check to see if $T^Z_{i}(P Z_j)=0$, noting that this involves computing the canonical form of $P Z_j$. If so, then we know that $P Z_j$ is of weight $i+1$, so we add it to $T^Z_{i+1}$ and continue. Every operator of weight $i + 1$ must be of this form for some $P$ and $j$ , so we are guaranteed to reach all operators of weight $i + 1$ this way.

Consequently, once we have finished the iteration, $T^Z_{i+1}(P)$ will contain all operators of weight exactly $i + 1$ modulo the code. We then set $T^Z_{i+1} (P) = 1$ for
all $P$ in $T^Z_{i}$ to ensure that $T^Z_{i}$ contains all operators of weight at most $i + 1$, which completes the induction.

Although this algorithm is not polynomial in time or space, it need only be evaluated once per code, and it does not take excessive time to evaluate it for many codes of reasonable size such as the aforementioned Golay code. It is to be hoped that it is practical to compute and use lookup tables for larger codes, though this will not scale indefinitely.

\section{Greedy circuit optimisation of the distance five circuit for $\nkd{17}{1}{5}$}
\label{sec:appendix-17-1-5-optim}

\begin{figure}[t]
\centering
\includegraphics[width=\columnwidth]{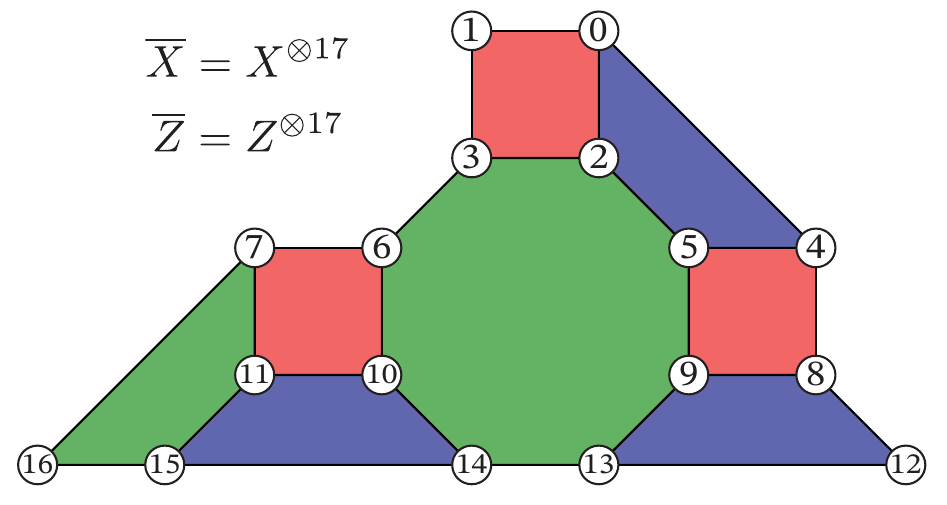}
\caption{Code definitions we use for the $\nkd{17}{1}{5}$ colour code. Each plaquette is associated with an $X$ and $Z$ type stabiliser acting on the qubits bordering the plaquette. We choose representatives of logical $\overline{X}$ and $\overline{Z}$ that act on all qubits.
\label{fig:app-17-1-5-code}}
\end{figure}

In \Cref{sec:demonstration} we use Comb to test the fault-tolerance of a set of magic state preparation circuits for the $\nkd{17}{1}{5}$ colour code, including two variants that are fault-tolerant to the distance of the code $d=5$. The second $d=5$ circuit is obtained from the first by a circuit optimisation procedure. This optimisation is made possible by the high speed that Comb is able to test candidate circuits.
In this appendix we describe that optimisation process in detail. 
The code definitions we adopt for $\nkd{17}{1}{5}$ are shown in \Cref{fig:app-17-1-5-code}.

As a starting point we begin with a circuit following the structure given in~\cite{chamberlandFaulttolerantMagicState2019}. This is a circuit composed of (a)~non-fault-tolerant preparation of $T \ket +$, (b)~flagged measurement of the logical stabiliser $SX$, (c)~full quantum error detection cycle. To make the circuit fault-tolerant (b) and (c) are repeated giving a final sequence (a)-(b)-(c)-(b)-(c). The quantum error detection cycle we employ measures each stabiliser generator of the code once, using a separate flag gadget for each. If any stabiliser measurement, or its flag, has a non-trivial outcome the circuit would be discarded. We first verify with Comb that this circuit is fault-tolerant, as expected.

The number of two-qubit gates in the circuit is reduced using a greedy approach that removes ancilla qubits.
These ancillas correspond to the measurement of a stabiliser eigenvalue or a flag on that measurement. An ancilla is selected and removed from the circuit along with any gates that touch it. After each removal we use Comb to test if the remaining circuit is still fault-tolerant. If it is, the circuit is accepted, else it is rejected and that ancilla remains in the circuit. In either case we proceed to test subsequent ancillas until we have tested all of them.

\begin{figure*} \caption{Fault tolerant preparation of $T \ket +$ in the $\nkd{17}{1}{5}$ colour code. This circuit is composed of a non-fault-tolerant state preparation, followed by a sequence of measurements of the logical stabiliser $SX$ and stabilisers of the $\nkd{17}{1}{5}$ code.} \label{fig:circuit}
\vspace{6pt}
\includegraphics[width=0.95\textwidth]{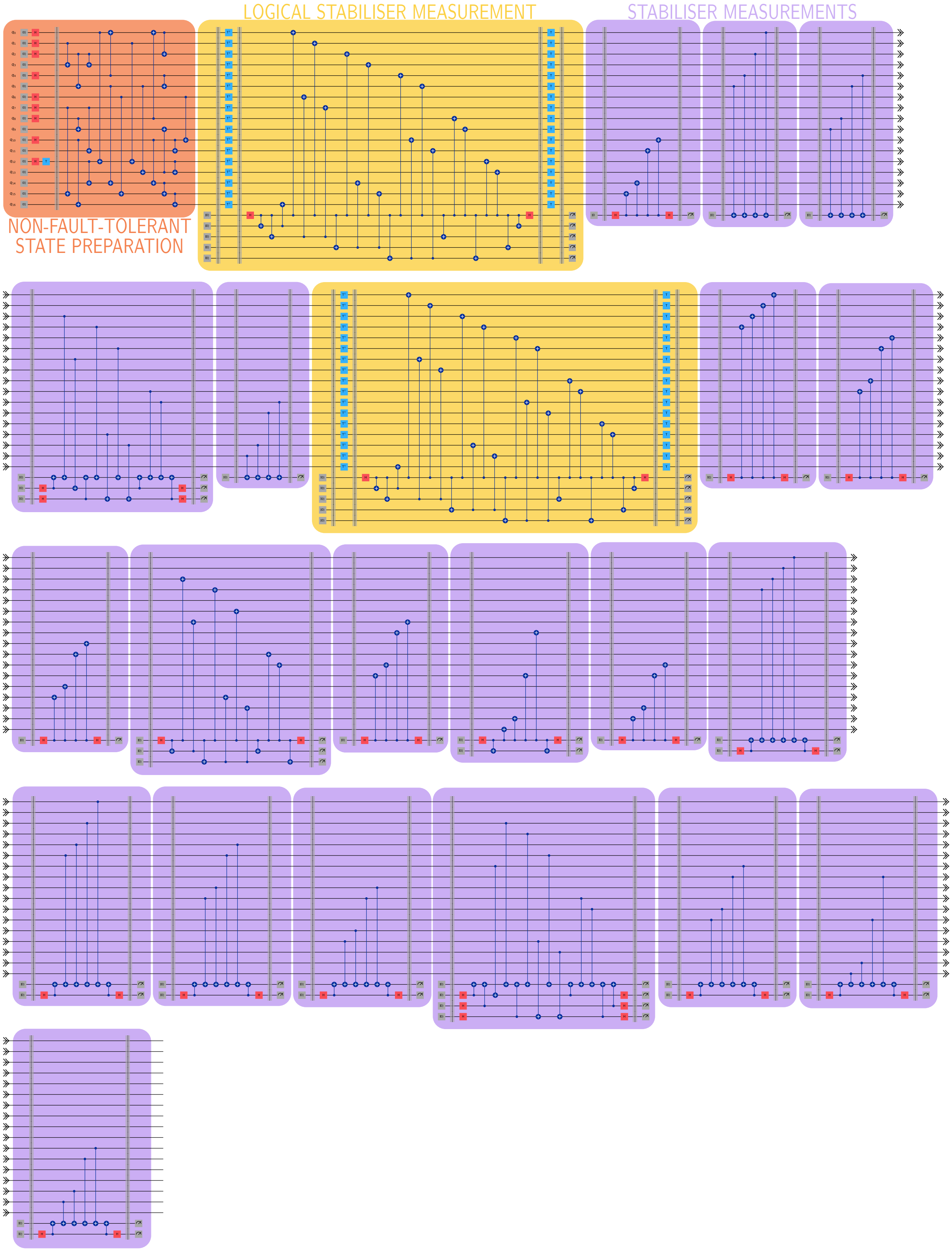}
\end{figure*}

Applying this procedure, we are able to reduce the number of two-qubit gates in the fault-tolerant $T \ket +$ magic state preparation for the $\nkd{17}{1}{5}$ colour code from 298 to 193. The final circuit we find is shown in \Cref{fig:circuit}.

\section{Independent verification of the $\nkd{17}{1}{5}$ circuits by simulation}
\label{sec:appendix-17-1-5-sims}

To independently test Comb's findings we simulate the magic state preparation circuits for the $\nkd{17}{1}{5}$ code using SymFT~\cite{fangSymFTUniversalFaultTolerant2026} and compute the scaling of their logical error rates with physical error rate. In this appendix we give full details of these simulations. The code definitions we adopt for $\nkd{17}{1}{5}$ are shown in \Cref{fig:app-17-1-5-code}.

\begin{figure*}[t]
\centering
\includegraphics[width=\textwidth]{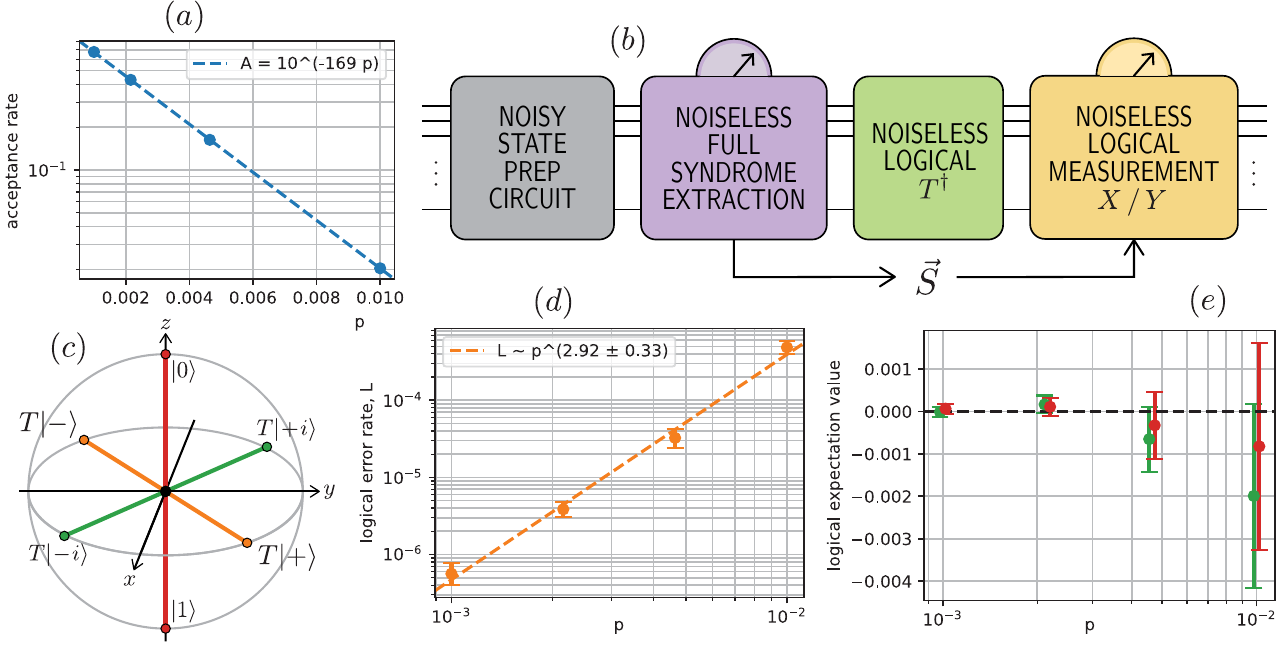}
\caption{
Simulation scheme outline and numerical results for the optimised $d=5$ circuit for the $\nkd{17}{1}{5}$ code. 
(a) Post-selection acceptance rate for the circuit. The acceptance rate drops exponentially with the physical error rate $p$.
(b) Simulation workflow executed in SymFT~\cite{fangSymFTUniversalFaultTolerant2026}. First, the magic state preparation circuit is simulated with circuit noise at rate $p$. Following this we noiselessly extract the full syndrome of the output state and apply a noiseless logical $T^\dagger$ rotation. 
Dependent on the syndrome we may now want to measure the logical $\overline{X}$ or $\overline{Y}$ operator in order to compute the logical error rate. 
To achieve this we repeat the simulation twice, once with a final measurement in the logical $\overline{X}$ basis and again in the logical $\overline{Y}$ basis.
(c) Logical error rate is computed from measuring the output state in the $\{ T|+\rangle, T|-\rangle \}$ basis. On a per-shot basis the syndrome determines whether measuring $\overline{X}$ or $\overline{Y}$ after the logical $T^\dagger$ corresponds to measuring in the $\{ T|+\rangle, T|-\rangle \}$ basis or the rotated $\{ T|+\!i\rangle, T|-\!i\rangle \}$ basis. To fully characterise the output state we can additionally measure in logical $\overline{Z}$. These three sets of measurement basis are indicated on the logical Bloch sphere.
(d) Logical error rate $L$ as a function of physical error rate $p$ showing scaling $L \sim p^3$ consistent with the circuit being fault-tolerant to $d=5$. Logical error rate is computed as the fraction of shots that return $T|-\rangle$ during the measurement in the $\{ T|+\rangle, T|-\rangle \}$ basis.
(e) Expectation value of the measurements in the two other basis, we see these are consistent with zero as expected.
\label{fig:app-numerics}}
\end{figure*}

The simulation scheme we use is sketched in \Cref{fig:app-numerics}(b). It begins by simulating the state preparation cirucit with circuit level noise. The circuit noise makes all operations faulty with the same physical error rate $p$: that is we insert bit-flip qubit initialisation and measurement errors, as well as single qubit and two qubit depolarising channels on gates, all with rate $p$. Following this we apply a sequence of noiseless measurements and rotations that allows us to compute the logical fidelity.

The logical fidelity is the overlap between the output state of the circuit and the ideal state $T|+\rangle$. Characterising the logical fidelity by reconstructing the logical Bloch vector $( \, \langle\overline{X}\rangle, \, \langle\overline{Y}\rangle, \, \langle\overline{Z}\rangle \, )$ and computing its projection onto the $T|+\rangle$ state would require a very large number of samples: $\sim\!10^{14}$ for logical error rates $\sim\!10^{-7}$. To avoid this sampling cost we measure directly in the $\{ T|+\rangle, T|-\rangle \}$ basis, \Cref{fig:app-numerics}(c).

Measuring in the $\{ T|+\rangle, T|-\rangle \}$ basis is achieved by adding the following noiseless operations to the end of the simulation, \Cref{fig:app-numerics}(b): (1) extracting the full syndrome $\vec{S}$, (2) applying a logical $T^\dagger$ rotation and (3) measuring in either the logical $\overline{X}$ or $\overline{Y}$ basis as determined by $\vec{S}$. 
We decode the syndrome $\vec{S}$ with a \emph{code-capacity decoder} (described below), this yields a physical Pauli error $E$ that we believe to be equivalent, up to stabilisers, to the error that is present on the state. 
If we encounter a syndrome that is not in the code-capacity decoder we reject the sample as an additional post-selection.
If $E$ commutes with the logical $\overline{Z}$ then the logical $T^\dagger$ acts correctly and $T|+\rangle$ is rotated back to $|+\rangle$ and so we measure in the $\overline{X}$ basis. However, if $E$ anti-commutes with $\overline{Z}$ then $T^\dagger$ is flipped to $T$ and $T|+\rangle$ is rotated instead to $|\!+\!i\rangle$, in this case we measure in the $\overline{Y}$ basis. To avoid the need to dynamically choose the final measurement basis at runtime we run the simulation twice: once with the measurement in the $\overline{X}$ basis and again in the $\overline{Y}$ basis. When we post-process the data we use the syndrome of each sample to determine whether its measurement corresponds to the $\{ T|+\rangle, T|-\rangle \}$ basis or the rotated $\{ T|\!+\!i\rangle, T|\!-\!i\rangle \}$ basis and aggregate the statistics accordingly.

The code capacity decoder we use for the final noiseless decoding step is constructed as a lookup table over all the correctable Pauli errors of the code. The set of all correctable Pauli errors is taken to be all Paulis with \emph{CSS weight} $\leq 2$, meaning that if we split them $P = P^X P^Z$ as is \Cref{preliminaries:pauli-decomposition} we require the weight (the number of non-identity elements) of both $P^X$ and $P^Z$ to be $\leq 2$. To give a clarifying example the CSS weight of $X_1 X_4 X_5$ would be 3 while $X_1 Z_4 X_5$ would only be 2. We iterate over all correctable errors and calculate their syndrome. If the syndrome is not already in the lookup table we add it, else we do nothing.

The logical error rate is the number of samples where we measure $T|-\rangle$ when measuring in the $\{ T|+\rangle, T|-\rangle \}$ basis. This is plotted versus physical error rate $p$ for the optimised $d=5$ circuit in \Cref{fig:app-numerics}(d), and for the full set of $\nkd{17}{1}{5}$ circuits in \Cref{fig:numerics}. Error bars shown on the logical error rates are 95\% Wilson confidence intervals. We plot the measurements in the two complementary basis of the optimised $d=5$ circuit in \Cref{fig:app-numerics}(e), with error bars showing the 95\% Wald confidence intervals. Finally the acceptance rates of post-selection are plotted for the optimised $d=5$ circuit in \Cref{fig:app-numerics}(a), and for all circuits in \Cref{fig:numerics}. 

We simulate the circuits at error rates $p=10^{-3}$, $10^{-2.666}$, $10^{-2.333}$, $10^{-2}$ with the following numbers of samples: for the non-FT and $d=3$ circuits $10^7$, $10^6$, $10^6$, $10^6$; for the $d=5$ circuits $10^8$, $5\times10^7$, $10^7$, $10^7$. The most expensive simulation, the $d=5$ unoptimised circuit at $p=10^{-3}$, took 63 minutes on an Apple Macbook Pro with Apple M4 processor.

\end{document}